# Resonant transport in a highly conducting single molecular junction via metal-metal covalent bond

Biswajit Pabi[1], Štepán Marek[2], Adwitiya Pal[3], Puja Kumari[4], Soumya Jyoti Ray[4], Arunabha Thakur[3], Richard Korytár[2], and Atindra Nath Pal[1*]

[1]Department of Condensed Matter and Materials Physics, S. N. Bose National Centre for Basic Sciences, Sector III, Block JD, Salt Lake, Kolkata 700106, India.
[2]Department of Condensed Matter Physics, Faculty of Mathematics and Physics, Charles University, 121 16, Prague 2, Czech Republic
[3]Department of Chemistry, Jadavpur University, Kolkata-700032, India.
[4]Department of Physics, Indian Institute of Technology Patna, Bihar- 801106, India.

**Abstract:**

Achieving highly transmitting molecular junctions through resonant transport at low bias is key to the next-generation low-power molecular devices. Although, resonant transport in molecular junctions was observed by connecting a molecule between the metal electrodes via chemical anchors by applying a high source-drain bias (> 1V), the conductance was limited to < 0.1 $G_0$, $G_0$ being the quantum of conductance. Here, we report electronic transport measurements by directly connecting a Ferrocene molecule between Au electrodes at the ambient condition in a mechanically controllable break junction setup (MCBJ), revealing a conductance peak at ~ 0.2 $G_0$ in the conductance histogram. A similar experiment was repeated for Ferrocene terminated with amine (-$NH_2$) and cyano (-CN) anchors, where conductance histograms exhibit an extended low conductance feature including the sharp high conductance peak, similar to pristine ferrocene. Statistical analysis of the data along with density functional theory-based transport calculation suggests the possible molecular conformation with a strong hybridization between the Au electrodes and Fe atom of Ferrocene molecule is responsible for a near-perfect transmission in the vicinity of the Fermi energy, leading to the resonant transport at a small applied bias (< 0.5V). Moreover, calculations including Van der Waals/dispersion corrections reveal a covalent like organometallic bonding between Au and the central Fe atom of Ferrocene, having bond energies of ~ 660 meV. Overall, our study not only demonstrates the realization of an air-stable highly transmitting molecular junction, but also provides an important insight about the nature of chemical bonding at the metal/organo-metallic interface.

**1. Introduction:**

The central motivation of molecular electronics is to reduce the size of the conventional electronic components by taking advantage of the quantum mechanical phenomena at the atomic scale, along with the structural versatility of molecules. By adopting the revolutionary idea proposed by Aviram and Ratner[1], various experimental techniques[2–19] were developed to create single-molecule junctions, allowing access to electrical[20–24], chemical[25–29], mechanical[30–32], magnetic[33], thermal[34–38] and optical[39,40] properties. Typically, various anchoring groups such as thiol (**-SH**)[11,41,42], amine (**-$NH_2$**)[43,44], pyridyl (**-PY**)[11], nitrile (**-CN**)[45], isonitrile (**-NC**)[46,47], carboxylic acid (**-COOH**)[48], nitro (**-$NO_2$**)[49], trimethyltin (**-$SnMe_3$**)[50], fullerene(**C60**)[51,52] are used to connect organic molecules to the metal electrodes to achieve physically stable structure with reliable electronic coupling at the interface. Since the conductance of a single molecule junction is primarily sensitive to the degree of orbital hybridization and energy-offset between the frontier molecular orbitals and the Fermi level of the metal electrodes[53,54], the presence of these anchoring group adversely affects the conductance in most of the cases[55]. With the goal of achieving high conductance, molecular junctions were realized either via direct metal-carbon coupling through molecules like oligocene[56], benzene[57–60], fullerene[54] or through direct Au-C covalent bonding in trimethyl tin (SnMe$_3$)-terminated polymethylene chains[50]. However, at ambient condition, the conductance remains far from the quantum of conductance ($G_0 = 2e^2/h \approx 77.6$ μS) mostly due to the off-resonant transport mechanism[50,58,61] either through coherent tunneling or via hopping as a result of large energy off-set.

Recently, there are efforts to access the resonant transport regime in biphenyl-porphyrin oligomers (bp-ppo)[62], donor-acceptor oligomers based on DPP[23] and amine or thiomethyl terminated oligophenyls[63]. Notably, the resonance could be achieved only by applying a high source drain bias (> 1V) or through a gate voltage in a three-terminal configuration[64] at cryogenic temperature. At such a high bias the molecular junction could become unstable due to the Joule heating, or, the characteristics of the junction can be modified by the chemical reaction occurring due to the higher current density[65,66]. Also, considering the realistic difficulties in fabricating three terminal single molecular devices, it is indeed desirable to reduce the energy off-set by exploring other possibilities of chemical functionalization.

Here, we focus on ferrocene, a chemically stable molecule at ambient condition, as the primary component of a single molecular device[67–73]. The unique structure of metallocene having a central transition metal at the center was shown to provide near-ballistic spin transport while incorporated between silver electrodes at cryogenic environment[74]. Despite recent reports on ferrocene or its derivative in molecular junctions[69–71,75–77], the detailed transport mechanism of ferrocene based junction is still elusive. In this study, we demonstrate the charge transport characteristics of a single Au/ferrocene/Au junction via direct metal-molecule coupling, exhibiting a significantly high conductance (~ one quarter of the quantum of conductance) at ambient condition. Density functional theory-based transport calculations suggest that the frontier molecular orbitals (LUMO) lies in close proximity to the metallic Fermi level of the electrodes for a certain orientation of the molecule, providing resonant transport at small applied bias (<0.5V). Further experiment along with the theoretical calculation, considering ferrocene terminated with two anchoring groups (**-NH$_2$**) and (**-CN**), display the similar high conductance conformation in addition to the features at the low conductance value, arising due to bonding with the chemical anchors. In general, our study reveals a way to obtain the resonant transport at the low bias regime along with a deeper understanding on the role of the spatial orientation of organometallic molecules on the charge transport mechanism.

## 2. Results & Discussions:
### 2.1. Conductance characterization of ferrocene single molecular junction
Typical conductance-distance breaking traces of ferrocene (FC) single molecular junction, connected to Au electrodes, is displayed in the left panel of figure 1b. Conductance plateau at 1 G$_0$ (G$_0$ = 2e$^2$/h being the quantum of conductance) demonstrates the formation of single atomic contact, i.e., only one Au atom in the narrowest cross section. However, upon further stretching, the conductance has a sudden fall due to the breaking of atomic contact followed by an opening a nanogap due to the snapback effect[78,79], leading to additional features below 1.0 G$_0$ when placed in a molecular environment. In this case, almost flat conductance plateaus can be observed in the region ~ 0.1-0.4 $G_0$ (figure 1b, left panel) following the post-rupture quantum tunneling features which confirm the successful formation of molecular junctions. Due to the pronounced thermal noise at room temperature, molecular plateaus are not always identical and vary from trace to trace. Thus, to get meaningful conductance value from statistically independent junction configurations, several thousands of traces are analyzed and represented as conductance histogram (1D histogram) in the middle panel of figure 1b (for bias voltage, $V_{bias}$ = 400 mV). Histogram is constructed using 35 bins per decade from 1800 molecular traces. Large counts at or around 1.0 $G_0$ appears due to repetitive formation of Au-atomic junction[80]. However, a well-defined conductance peak with maximum at $(2.20 \pm 0.03) \times 10^{-1}$ $G_0$ is also evident from the histogram, and it corroborates the most probable conductance value of the Au/ferrocene/Au single molecular junction at room temperature. Since it is technically unachievable to obtain any vibrational fingerprint of the molecule in a notched-wire break junction set up using the conventional techniques like surface enhanced Raman spectroscopy (SERS), atomic force microscopy (AFM) or inelastic electronic spectroscopy (IETS) at room temperature, we performed controlled experiment with DCM and found the peak to be completely absent, as shown in Supplementary Information, figure S5[81]. The bias dependent measurement shows an enhancement of the conductance peak by almost a factor of 1.5 when the bias is increased from 20 mV to 400 mV (inset of figure 1b, middle panel; also see Supplementary Information, figure S6 for the histograms at 20 mV and 50 mV bias). To get more insight about the evolution of the molecular junction during stretching, we constructed a 2D conductance-distance density plot in a semi logarithmic scale. The distance or electrode separation was calibrated following the procedure using tunneling experiments[82]. Right panel of the figure 1b shows the typical 2D histogram of FC, which is generated by compiling the same traces used in the 1D histogram with 80 bins per decade. High density data clouds at or above $1G_0$, represents the Au atomic junction. A second horizontal data cloud (marked with a black dashed circle) is noticed from $3.00 \times 10^{-1}$ $G_0$ to $8.00 \times 10^{-2}$ $G_0$ and interestingly, conductance of the molecular plateaus does not change significantly with the increase of electrode separation, similar to the silver-vanadocene junction reported by Pal et al[74].

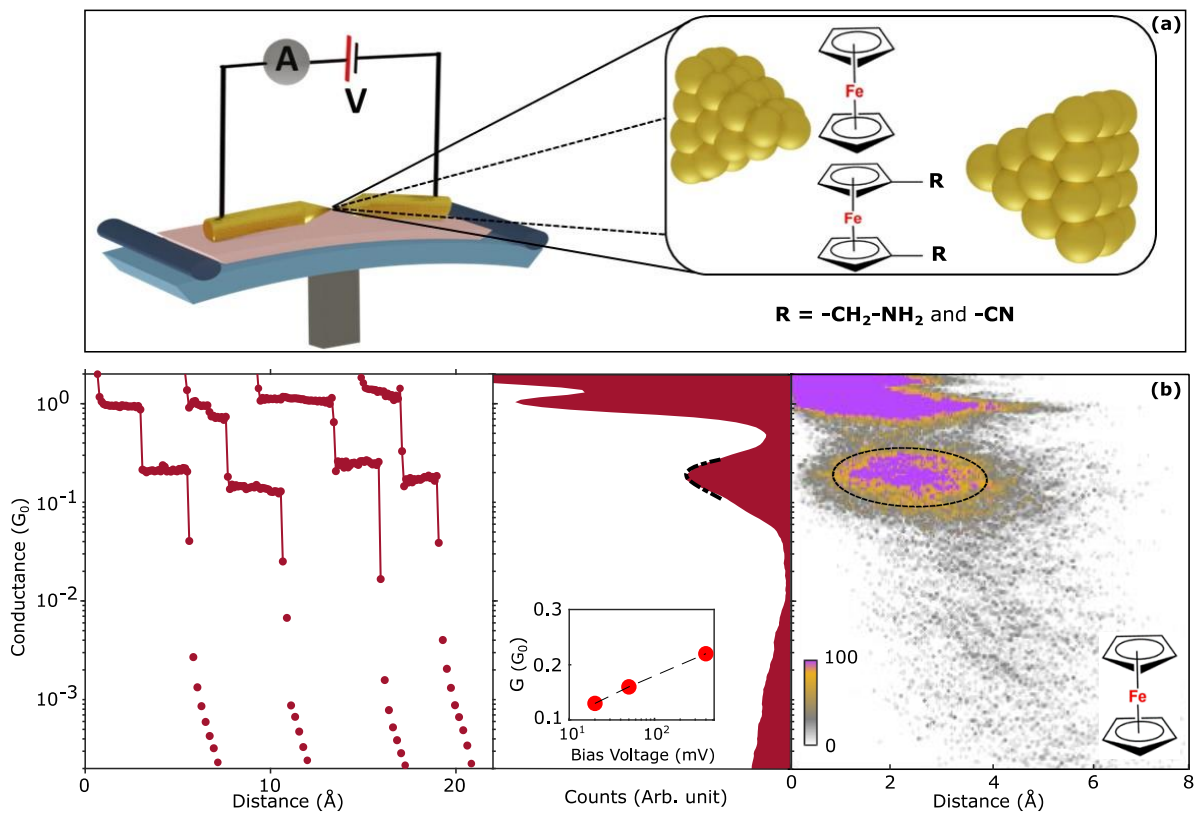

*Figure 1: (a) Schematic illustration of MCBJ set up along with the molecular structure (ferrocene, 1,1'-bis(aminomethyl)ferrocene and 1,1'-dicyanoferrocene) used here. (b) Left Panel: Typical breaking conductance-distance traces of Au/ferrocene/Au molecular junctions in semi-logarithmic scale, with traces shifted horizontally for clarity. Middle Panel: One dimensional conductance histogram, constructed from 1800 molecular traces using 35 bins per decade. Black dash-dotted line represents the Gaussian fitting of the corresponding molecular peak with the most probable value $(2.20 \pm 0.03) \times 10^{-1}$ $G_0$. Inset: conductance of the junction as a function of applied bias voltage. Right Panel: Conductance-distance density plot of ferrocene molecular junctions based on the same 1800 molecular traces, constructed using logarithmic binning (50 bins per decade), where black circle shows the molecular plateaus. Inset: chemical structure of ferrocene.*

## 2.2. Formation mechanism and correlation analysis

The conductance histogram is now considered using traces with and without the molecular features (shown in the Supplementary Information, figure S7). Histogram of the molecular traces reveals a shallow peak at ~1.2 $G_0$ in addition to the atomic peak at ~ 1.0 $G_0$, arising possibly due to the formation of precursor configuration prior to the atomic junction in line with the previous reports [83,84]. To get a clearer picture about the evolution of the junction, we divide the molecular traces into three different sub-categories based on the presence of plateaus at $G_p$, $G_{Au}$ and $G_m$ (Case-I), at $G_p$ and $G_m$ (Case-II), at $G_{Au}$ and $G_m$ (Case-III) (conductance values of $G_p$~1.2 $G_0$, $G_{Au}$~1.0 $G_0$ and $G_m$~0.2 $G_0$). Characteristic traces of three different subcategories are shown in figures 2a-c with the inset revealing the possible junction geometries, where L and R denote the left and right electrodes with M as molecule. Figure 2d shows the conditional conductance histogram of the three categories and the pie chart (inset) showing the probability to find the different cases: Case-I [20%], Case-II [13%] and case-III [67%]. From the above analysis, it is obvious that precursor configuration is not a necessary condition to have the molecular junction. In fact, the most possible scenario [Case-III] does not even involve the precursor configuration. Similar precursor configuration was observed for diatomic[83,85] and organometallic[84] molecule-based junction which primarily occurs when the atomic contact lies in parallel to the molecular junction (shown schematically in the insets of figures 2a and 2b). As a next step, we explore the statistical relationship of these configurations with the help of two dimensional cross-correlation histogram (2DCH) technique, introduced in reference [86] (also see Supplementary Information, section 2d). Using this technique, a map is generated for arbitrary conductance pairs $G_i$ & $G_j$ where the color-scale indicates the sign of the correlation among the conductance pairs (orange and blue represents the positive and negative correlation respectively). Figure 2e shows the 2D cross correlation map of ferrocene (FC) molecular junctions where horizontal and vertical axes correspond to the two conductance, $G_i$ & $G_j$. Now we consider three different conductance regions: (a) R1- molecular conductance (0.09 - 0.40 $G_0$); (b) R2 - single atomic conductance (0.80 - 1.05 $G_0$) and (c) R3 – precursor conductance (1.05 - 1.40 $G_0$). In line with the conditional analysis, a positive correlation between the precursor peak (R3) and the molecular peak (R1) is noticed, marked with the black rectangle in figure 2e. However, the correlation between R1 and R2 is negative, marked by the white rectangle, similar to the observation of Ag-CO-Ag junction[83]. This observed negative correlation is not an anti-correlation in the presence of these two configurations. Rather, the presence of one plateau restricts the occurrence of other in longer length, as evident from the conditional analysis shown in the Supplementary Information, figure S8[86]. Overall, this combined analysis reestablishes the eventual formation of the precursor configuration in which molecule binds alongside to the atomic contact and opens additional channel which results in an increase of the conductance up to ~ 1.3 $G_0$.

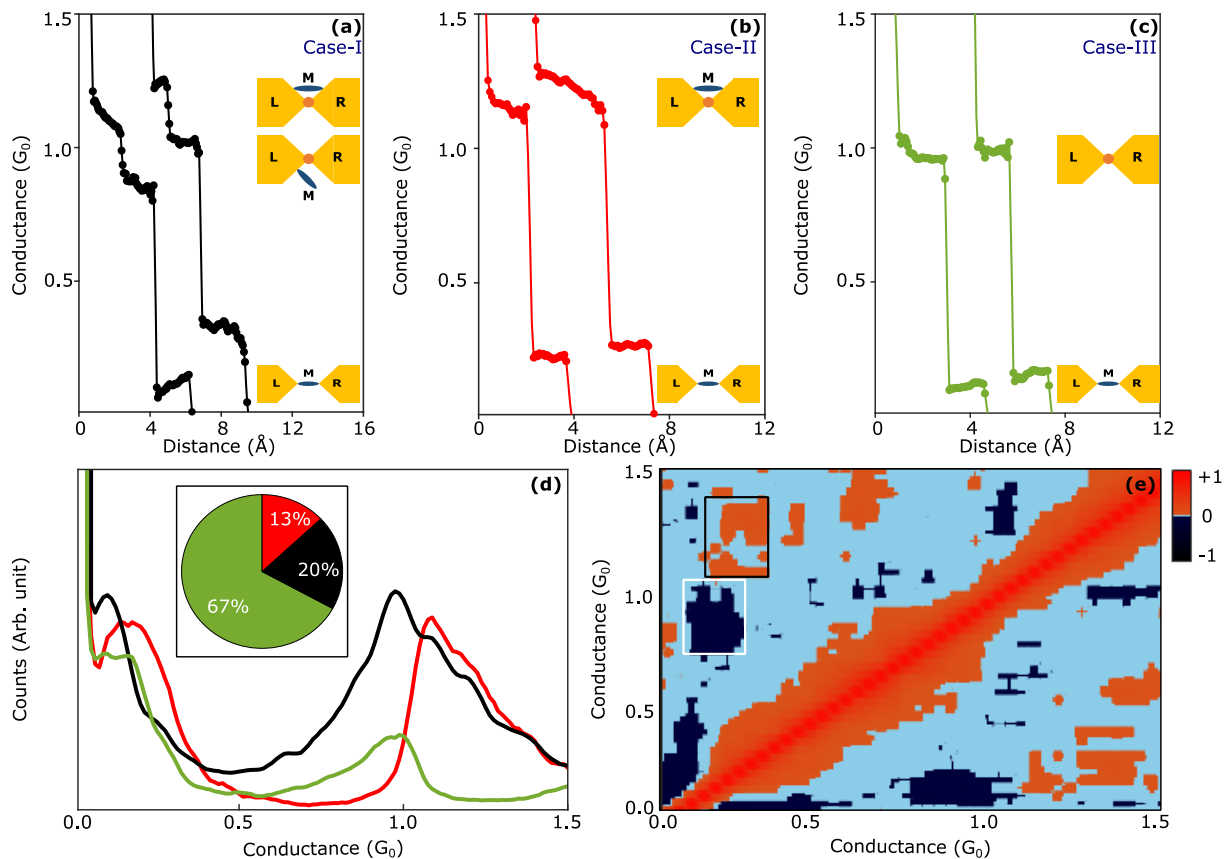

*Figure 2:* (a - c) Sets of conductance traces with different combinations of plateaus during breaking process: 1. Case-I: Precursor, $G_p$ ~1.2 $G_0$, Atomic, $G_{Au}$ ~1.0 $G_0$ and Molecular, $G_m$ ~0.2 $G_0$ (Black), 2. Case-II: Precursor and Molecular (Red), 3. Case-III: Atomic and Molecular (Green). Inset: schematic representation of the possible junction geometries where L, R and M denote the left electrode, right electrode, and molecule in between, respectively. (d) Conditional conductance histogram in linear scale, considering traces of case-I (Black), II (Red) and III (Green) using 200 bins. Inset: Pie chart of percentage of traces contributing different cases. (e) 2D correlation map of ferrocene molecular junctions for breaking conductance traces, where negative correlation (white rectangle) between the molecular region and atomic peak at 1$G_0$ is clearly visible. Conductance of single-molecule configuration and precursor configuration is positively correlated (black rectangle).

## 2.3. Effect of anchoring groups

Now we focus on the influence of anchoring group on the conductance behavior of ferrocene molecular junction. Left panels of figures 3a and 3b, respectively, show the conductance-distance breaking traces of amine-terminated ferrocene (1,1′-bis(aminomethyl)ferrocene, abbreviated as FC-NH$_2$) and cyano-terminated ferrocene (1,1′-dicyanoferrocene, abbreviated as FC-CN) single molecular junctions connected to Au electrodes. In contrast to FC, molecular plateaus of FC-NH$_2$ and FC-CN have a broad range of conductance values and subsequent plateaus can also appear in a single trace[71]. Transitions among these plateaus are not always sequential, their order completely arbitrary. Thus, introduction of anchoring group provides more conductance plateaus which signifies the multiple junction geometries. One-dimensional conductance histograms of FC-NH$_2$ and FC-CN molecular junctions are shown in the middle panels of figures 3a and 3b. Histograms are constructed from 5000 molecular traces using 35 bins per decade. Closely looking into the histogram, well defined conductance peaks at $(3.57 \pm 0.09) \times 10^{-1}$ G$_0$ and $(2.47 \pm 0.03) \times 10^{-1}$ G$_0$ are observed for FC-NH$_2$ and FC-CN junctions. These conductance values agree well within a factor of two of the conductance peak of pristine FC and from now on, we address the high conductance peak as **H**. Additionally, FC-NH$_2$ and FC-CN reveal a rather broad conductance feature beneath the high conductance peak (**H**) which can be convoluted into two individual conductance peaks with maxima at $(1.21 \pm 0.11) \times 10^{-2}$ G$_0$ and $(2.99 \pm 0.35) \times 10^{-3}$ G$_0$ for FC-NH$_2$ and at $(1.57 \pm 0.09) \times 10^{-2}$ G$_0$ and $(5.19 \pm 0.22) \times 10^{-3}$ G$_0$ for FC-CN. We refer to these conductance regions as **M** and **L** in descending values of conductance. **M** and **L** configurations show a trend: FC-CN > FC-NH$_2$ and their conductance values are distributed over two orders of magnitude between ~ $10^{-2}$ to $10^{-4}$ G$_0$ which hint at the structural variety of geometries due to the easy rotation of FC's cyclopentadiene rings[87]. All the conductance values are summarized in Table 1. Representative density plots of FC-NH$_2$ and FC-CN (right panel of figures 3a and 3b respectively) are constructed from the same 5000 traces using 80 bins per decade. A distribution of molecular plateaus over a broad range (as evident from the traces in the left panels of figures 3a and 3b ), along with the slanted data cloud  (see right panels of figures 3a and 3b) establishes multiple stable configurations appearing due to the chemical anchors[81].

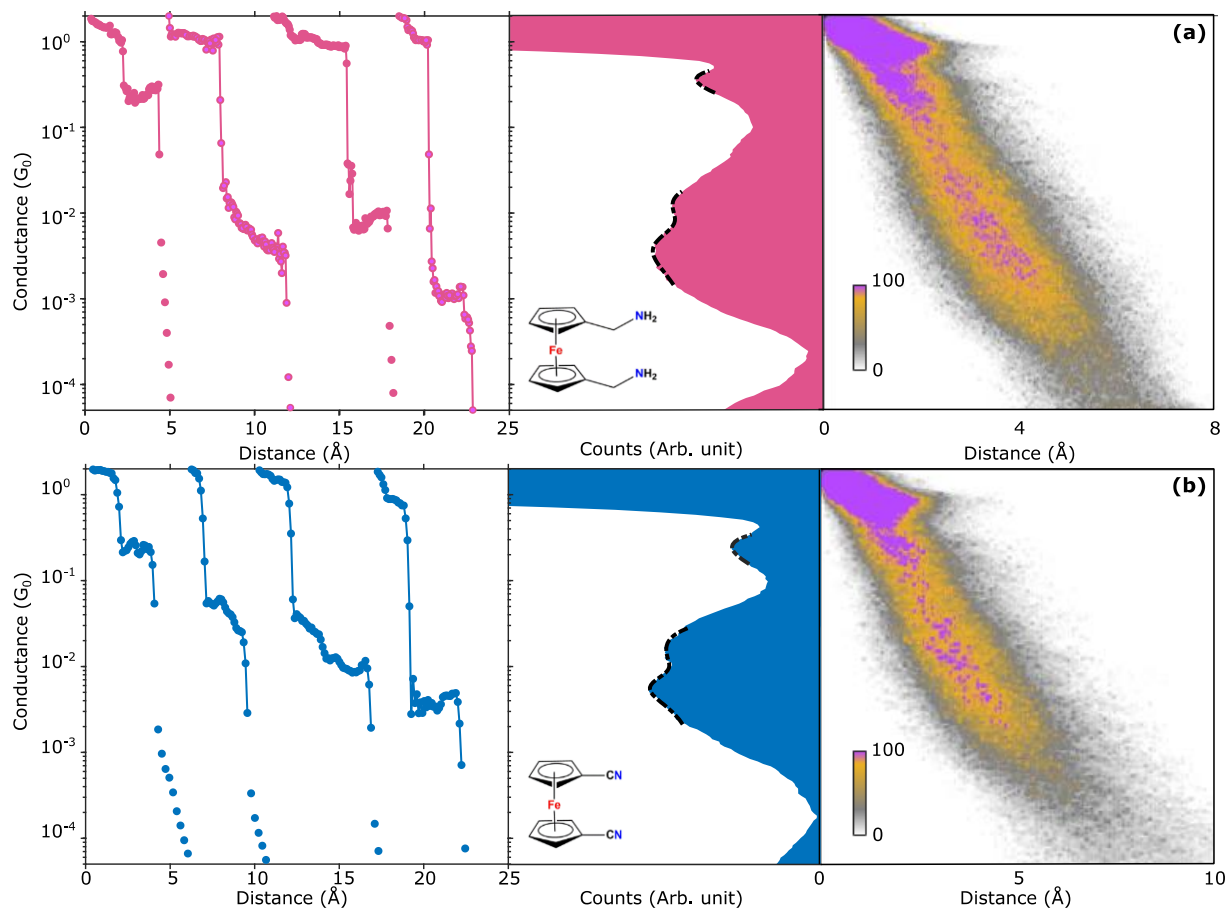

*Figure 3:* (a) Left Panel: Representative breaking conductance traces of Au/1,1'-bis(aminomethyl)ferrocene/Au molecular junction in a semi logarithmic scale, with traces shifted horizontally for clarity. Middle panel: 1D logarithmic conductance histogram, constructed by analyzing 5000 individual molecular traces using 35 bins per decade. Black dash-dotted line depicts the Gaussian fitting to the corresponding conductance peak and the inset shows the molecular structure. Right Panel: 2D conductance- distance density plot, generated from the same 5000 conductance traces using 80 bins per decade. The zero distance point is assigned here at 2G$_0$. (b) Similar characterization of Au/1,1'-dicyanoferrocene/Au molecular junction.

*Table 1: Conductance parameters of single molecular junctions*

| Parameter | Molecule | Configuration | | |
|---|---|---|---|---|
| | | High G (**H**) | Mid G (**M**) | Low G (**L**) |
| *Conductance (G$_0$)* | FC | $(2.20 \pm 0.03) \times 10^{-1}$ | ------------- | ------------- |
| | FC-NH$_2$ | $(3.57 \pm 0.09) \times 10^{-1}$ | $(1.21 \pm 0.11) \times 10^{-2}$ | $(2.99 \pm 0.35) \times 10^{-3}$ |
| | FC-CN | $(2.47 \pm 0.03) \times 10^{-1}$ | $(1.57 \pm 0.09) \times 10^{-2}$ | $(5.19 \pm 0.22) \times 10^{-3}$ |
| *Most probable stretching length (Å)* | FC | $0.7825 \pm 0.0095$ | ------------- | ------------- |
| | FC-NH$_2$ | $0.7949 \pm 0.0045$ | $1.871 \pm 0.003$ | $3.012 \pm 0.009$ |
| | FC-CN | $0.8486 \pm 0.0069$ | $2.174 \pm 0.003$ | $4.407 \pm 0.017$ |
| *Slope, (Å$^{-1}$)* | FC | $-(2.87 \pm 0.75) \times 10^{-2}$ | ------------- | ------------- |
| | FC-NH$_2$ | $-(5.05 \pm 0.60) \times 10^{-2}$ | $-(4.13 \pm 0.30) \times 10^{-2}$ | $-(5.87 \pm 0.21) \times 10^{-2}$ |
| | FC-CN | $-(4.58 \pm 0.57) \times 10^{-2}$ | $-(5.03 \pm 0.54) \times 10^{-2}$ | $-(8.78 \pm 0.40) \times 10^{-2}$ |

**2.4. Stretching length histogram and average traces**

To compare the evolution of the junction during stretching for the three different types of molecules, characteristic stretching length histograms for each conductance regime (**H**, **M** or **L**) are constructed by considering length from the relative zero position at 0.5 G$_0$ to one order of magnitude below the most probable conductance peak (adopting the procedure mentioned in the references [88,89]). Figures 4a-c display the typical stretching length histogram of FC, FC-NH$_2$ and FC-CN for **H** (blue), **M** (red) and **L** (yellow) conductance regimes.  The most probable plateau length is determined by the Gaussian fitting of the distribution (shown by the black dashed line), also summarized in Table 1. Notably, a positive shift in the distribution of **M** and **L** configurations compared to the **H** configuration, suggests that they correspond to a relatively longer conformation. While the larger width of the distribution for **M** and **L** configurations (figures 4a-c) illustrates the widespread variation in geometry upon stretching the junction[6,88], larger stretching length indicates higher formation stability. Furthermore, to probe the conductance decay during the elongation of the junction, average conductance-distance traces are shown in figures 4d-f for three different molecular junctions in the conductance regimes **H**, **M** or **L**. Almost flat plateau of the high conductance (**H**) region resembles to the atom-like contact[80], compared to the low conductance plateaus which are possibly originating due to the chemical anchors.

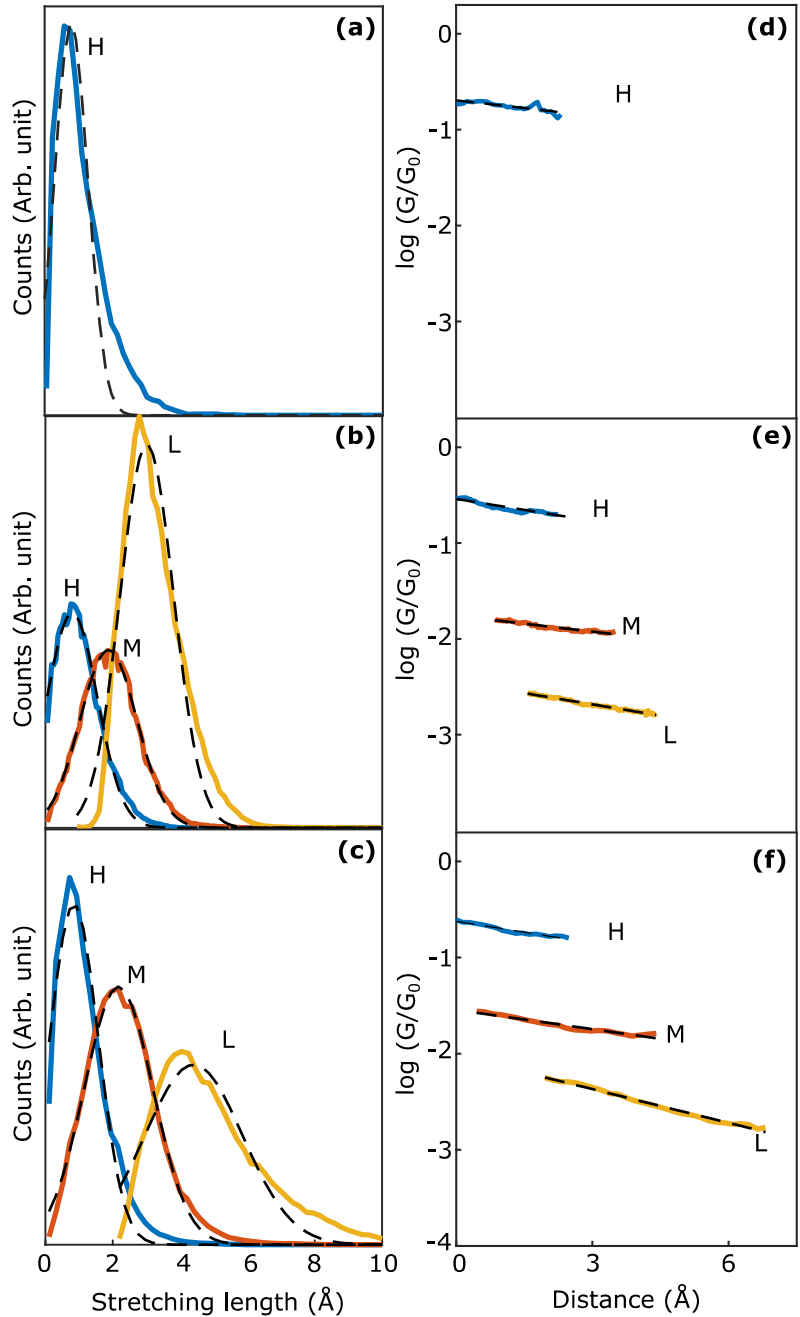

***Figure 4:*** *(a - c) Stretching length histogram of ferrocene, 1,1'-bis(aminomethyl)ferrocene and 1,1'-dicyanoferrocene for three different conductance regime (blue-H, red- M and yellow-L). Histogram of a particular conductance region is constructed by considering length from the relative zero position at 0.5 $G_0$ to one order of magnitude beneath the most probable conductance value of corresponding regime. Black dash line depicts the Gaussian fitting of the histogram which yields the most probable stretching length. (d - f) Statistically averaged conductance-distance traces of ferrocene, 1,1'-bis(aminomethyl)ferrocene and 1,1'-dicyanoferrocene . Here blue (red, yellow) color represents average traces of the high conductance regime, H (M, L). Black*

*dash line represents the linear fitting of the corresponding average traces to calculate the slope,* $\frac{\partial(\log G/G_0)}{\partial(\Delta x)}$

## 2.5. Theoretical calculations and interpretation of experimental results
### 2.5.1. High conductance feature

To gain insight into the origin of measured conductance, we now compare experimental results with detailed DFT - based calculations. First, we consider two possible geometries (parallel and perpendicular orientation of FC with respect to the electrodes) and for each geometry, energies at different distances of the electrodes from the central Fe atom are evaluated. Example coordinate files are provided in the Supplementary Information, Section (j). The scale of the energies for different distances is large compared to the thermal energy accessible to the system (see Supplementary Information, figure S9) and hence, each geometry can be considered stable with respect to the thermal fluctuations. Furthermore, we have also investigated the nature of bonding between the Au-Fe and Au-C for perpendicular and parallel geometries, respectively. Note that the incorporation of van der Waals correction seems to have significant impact on the binding energies for both the geometries. The bond energy was found to be significantly higher, ~ -660 meV for perpendicular and ~ -1610 meV for the parallel geometries, suggesting formation of

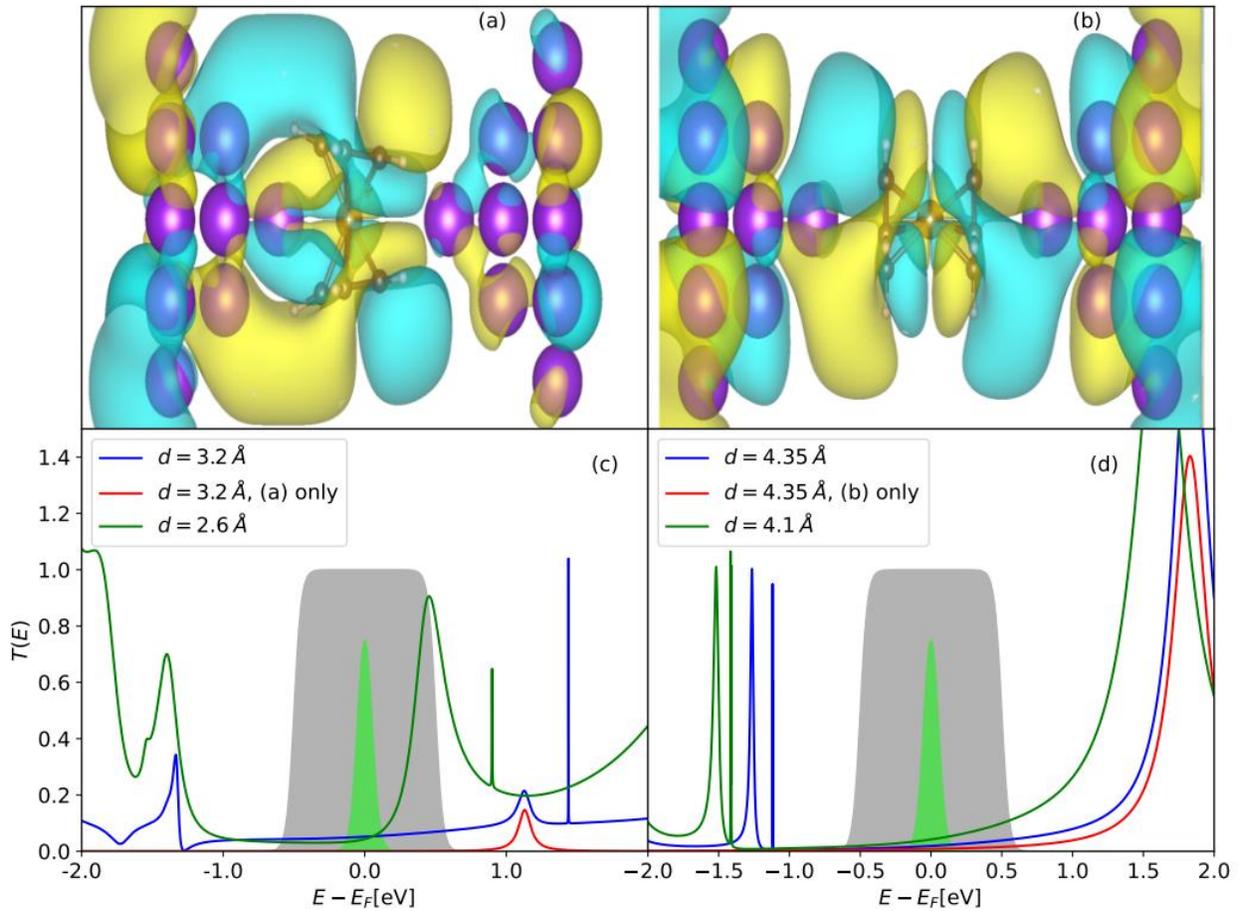

*Figure 5: (a-d) DFT calculations of electronic structure and transport in directly bound ferrocene. Top panels depict perpendicular (a) and parallel (b) geometries and scattering states at equal isosurface levels (yellow is positive, cyan is negative). Bottom panels show transmission functions*

*of the corresponding geometries (c-d). The transport windows for perpendicular (c) and parallel (d) geometries are given as filled curves for bias of 1 V (gray) and infinitesimal bias (green), both at 300 K. Transmission functions at different electrode distances d from center of the junction are shown, as well as transmission due to the displayed scattering state (see Supplementary Information for details).*

covalent bonding (see Supplementary Information, section 4f for details). Now, the transmissions for these two geometries are calculated and conductance (G) can be written as,[90]

$$G = \frac{I}{V} = \frac{2e^2}{h}\int_{-\infty}^{\infty} T(E) \frac{f_L\left(E;k_BT,E_F+\frac{eV}{2}\right) - f_R\left(E;k_BT,E_F-\frac{eV}{2}\right)}{eV} dE,$$

where $e$ is the elementary charge, $V$ is the bias of the junction and $f_{L/R}$ is the Fermi distribution of the left/right reservoir, at temperature $T = 300$ K. The distribution of the left and right reservoir is shifted symmetrically by the bias of the junction. The difference between the Fermi distributions is called the transport window. For conductance to be significant, the overlap of the transmission function and the transport window must be non-negligible. The transmission functions for both geometries are displayed in figures 5c-d. In perpendicular geometry, the transport window overlaps with a resonance for bias of about 1 V but does not overlap significantly for infinitesimal bias. The wavefunction responsible for this resonance (see figure 5a) resembles the LUMO orbital of gas-phase FC, shown in the Supplementary Information, figure S10. In the parallel geometry (figures 5b,d), the resonance of LUMO orbital is shifted further away from the Fermi energy and the transmission peak does not overlap with transport window even for bias of 1V. The resonance also corresponds to a peak in the local density of states on the iron atom, as described in the Supplementary Information, Section (g). The resonant transport in the perpendicular geometry yields relatively high conductance 0.1-0.3 $G_0$ and a stronger bias dependence in contrast to the off-resonant parallel geometry having conductance values in the range of 0.02-0.09 $G_0$. See Supplementary Information (Figure S11) for values of conductance at different bias for both the geometries. Comparing the measured conductance along with its strong bias dependence, we suggest that the perpendicular geometry dominates in the ensemble of junctions over the parallel geometry. The observation of high conductance peak in all three molecules along with similar stretching length (figures 4a-c) suggest that it originates primarily from the perpendicular geometry via Au-Fe bond. Additionally, we also explored the effect of ring rotation at a fixed geometry, but the calculated transmission function remains nearly unchanged, as shown in Supplementary Information, figure S12.

### 2.5.2. Low conductance feature

In this subsection, we explore the conductance behavior of ferrocene molecule coupled to Au electrode via -NH$_2$ and -CN anchoring groups. Possibility of binding through the chemical anchors of FC-NH$_2$ and FC-CN is clearly evident from the calculated negative binding energies for these two configurations (see Supplementary Information, Table S2) and corresponding coordinate files are provided in the Supplementary Information, Section (j). Figures 6c and 6f demonstrate the transmission curves for both the geometries when connected with the chemical anchors (termed as anchoring geometry), exhibiting transmission in the range of $10^{-2} - 10^{-3}$ in the vicinity of metallic Fermi level. Thus, coupling with the anchoring groups may provide multiple conformations of the molecule inside the junction, leading to the broad conductance features with larger stretching length and tilted plateaus. The scattering wave functions indicate that primarily HOMO – 1 orbital is responsible for electron transport of both FC-NH$_2$ (Figures 6a-b) and FC-CN junctions (Figures 6d-e).

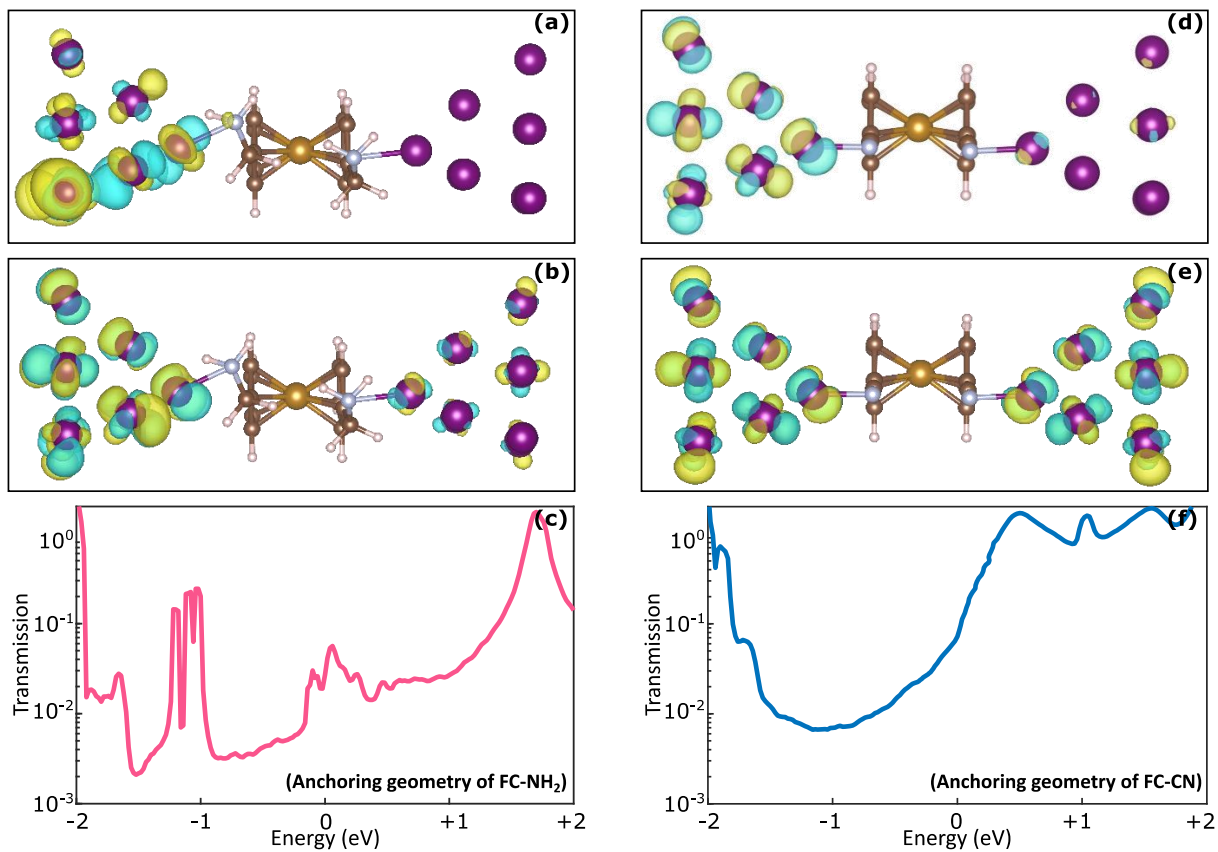

*Figure 6: (a-b) Transport states of anchoring geometry of Au/1,1'-bis(aminomethyl)ferrocene/Au : HOMO (a) and HOMO – 1 (b). Yellow isosurfaces have positive values. (c) Transmission curve of the same geometry as a function of energy under zero – bias. (d-f) similar characterization of Au/1,1'-dicyanoferrocene/Au junction. Isosurface levels of orbitals are chosen same for both geometries.*

### 3. Conclusions:

In conclusion, we have demonstrated the resonant transport at room temperature by connecting an organometallic molecule between Au electrodes. By measuring conductance-distance traces for thousands of statistically independent junctions using a mechanically controllable break junction set-up, we observe that pristine Ferrocene binds strongly with the Au electrodes and provides a significant high conductance (~ 0.1-0.3 $G_0$ in the bias range of 20 - 400 mV) in the conductance histogram. The density functional calculations suggest that both perpendicular and parallel orientations of the molecule with respect to electrodes are energetically favorable configurations via covalent-like bonding. Transmission curve for the perpendicular geometry reveals a sharp peak near metallic Fermi level which leads to a conductance value closed to the experimentally observed conductance as well as a stronger bias dependence. Whereas similar calculations in parallel geometry shows a conductance in the range of 0.01 – 0.09 $G_0$ and having a rather weak bias dependence. We also observed that the inherent rotation of cyclopentadienyl rings of Ferrocene has negligible impact on the transmission curves. Furthermore, to understand the effect of chemical anchors, we performed similar experiment on ferrocene terminated with two different anchoring groups (**-NH$_2$**) and (**-CN**). Both molecules display the high conductance peaks like pristine ferrocene in the conductance histogram, along with additional low conductance features. While the perpendicular geometry with Au facing the Fe atom might be responsible for the high conductance peak, the low conductance feature originate mainly due to the multiple binding geometries with the anchoring groups. Through this study we have not only explored the charge transport mechanism through a single organometallic molecule directly coupled to the metal electrodes via metal-metal chemical bond, but also provide a mechanism to exhibit long-range intramolecular transport which is important for next generation molecular devices working at ambient condition.

## 4. Methods:
### 4.1. Instrumentation for molecular characterization

$^1$H and $^{13}$C NMR spectra were obtained with BRUKER 300 MHz FT-NMR spectrometer and the chemical shifts are reported in ppm, using tetramethylsilane as an internal standard and were referenced to the residual solvent as follows: CDCl$_3$ = 7.26 ($^1$H), 77.16 ($^{13}$C) ppm at 22 ˚C. For $^1$H NMR, coupling constants *J* are given in Hz and the resonance multiplicity is described as s (singlet), br s (broad singlet), d (doublet), t (triplet), m (multiplet).

### 4.2. Preparation of ferrocene for conductivity studies

Commercially available ferrocene, purchased from LOBA Chemie, was recrystallized from dichloromethane (DCM) by the slow evaporation method at room temperature. These pure orange crystals were used directly for the conductance measurements.

### 4.3. Synthesis of 1,1′-bis(aminomethyl)ferrocene

1,1′-bis(aminomethyl)ferrocene was synthesized from 1,1′-ferrocenedimethanol according to literature reported procedure[91]. A solution of 1,1′-ferrocenedimethanol (932 mg, 3.78 mmol) in glacial acetic acid (10 ml) was charged with NaN$_3$ (2.95 g, 45.46 mmol) and stirred at 50 ˚C for 4h. The acid was neutralized with saturated NaHCO$_3$ solution and the reaction mixture was extracted with CH$_2$Cl$_2$ (3×20 ml). The organic layer was washed with brine, dried over Na$_2$SO$_4$, and taken to dryness. The crude residue of 1, 1′-bis(azidomethyl)ferrocene was purified by column chromatography in silica gel with hexane as the eluant to obtain pure 1,1′-bis(azidomethyl)ferrocene (984 mg, 3.32 mmol) as orange liquid. The 1,1′-bis(azidomethyl)ferrocene (984 mg, 3.32 mmol) was dissolved in EtOH (10 ml) and water (3 ml) and to this solution, NH$_4$Cl (810 mg, 15.1 mmol) and freshly activated Zn powder (594 mg, 9.08 mmol) were added. The suspension was stirred at room temperature for 24 h, after which the suspension was filtered and filtrate was taken to dryness. The residue was partitioned between EtOAc (3×20 ml) and 1 N aqueous ammonia (10 ml) and the organic layer separated, washed with brine, and dried over Na$_2$SO$_4$. The solvent was removed under reduced pressure and the residue was purified by column chromatography in basic alumina with methanol as the eluant. The resulting pure 1,1′-bis(aminomethyl)ferrocene (795 mg, 3.26 mmol) was obtained as dark yellow oil. $^1$H NMR (CDCl$_3$, 300 MHz) $\delta$ = 4.13 (t, 4H, *H*$_{Fc}$, *J* = 3 Hz), 4.09 (t, 4H, *H*$_{Fc}$, *J* = 3 Hz), 3.55 (s, 4H, C*H$_2$*), 1.76 (br s, 4H, N*H$_2$*). $^{13}$C NMR (CDCl$_3$, 75 MHz) $\delta$ = 90.9, 68.1, 67.5, 41.2.

## 4.4. Synthesis of 1,1′-dicyanoferrocene

1,1′-dicyanoferrocene was synthesized from ferrocene according to literature reported procedure[92]. 1,1′-diformylferrocene (709 mg, 2.92 mmol), $NH_2OH \cdot HCl$ (529 mg, 7.61 mmol), KI (962 mg, 5.8 mmol) and ZnO (472 mg, 5.8 mmol) were suspended in 15 ml of $CH_3CN$ and stirred for 48 h at 95 °C. The reaction mixture was cooled to room temperature and 10 ml aqueous solution of 5% $Na_2S_2O_3$ was added in a single portion and the reaction mixture was stirred for another 15 min. Then the reaction mixture was filtered through a filter paper and the filtrate extracted with $CH_2Cl_2$ (2×15 ml). The combined organic layers were dried over $Na_2SO_4$ and the solvent was removed under reduced pressure. The residue was purified by column chromatography in neutral alumina with EtOAc/*n*-hexane (1/4, *v/v*) as the eluant to obtain pure 1,1′-dicyanoferrocene (613 mg, 2.6 mmol) as orange powder. $^1$H NMR ($CDCl_3$, 300 MHz) $\delta$ = 4.82 (t, 4H, $H_{Fc}$, $J$ = 3 Hz), 4.60 (t, 4H, $H_{Fc}$, $J$ = 3 Hz). $^{13}$C NMR ($CDCl_3$, 75 MHz) $\delta$ = 118.3, 73.9, 73.5, 54.7.

## 4.5. Conductance measurements

We choose MCBJ technique to measure the electrical conductance of three organometallic molecular junction, illustrated in figure 1a (abbreviated as: FC – ferrocene, FC-$NH_2$ – 1,1′-bis(aminomethyl)ferrocene, FC-CN – 1,1′-dicyanoferrocene), at room temperature and ambient condition[93,94]. A notch is first created on an Au wire (0.1 mm diameter, 99.998%, Alfa aesar) and placed on top of a flexible substrate (Phosphor Bronze) using two component epoxy glue (Stycast 2850 FT with catalyst 9), schematically shown in figure 1a. Upon bending the substrate, the wire is broken from its weak spot creating a sub-nanometer gap which can be controlled using a piezo electric actuator with sub-angstrom precision. Molecules can then be easily trapped between the leads and this process is repeated several times to address many independent configurations. Throughout this breaking/making process, electrical current through the junction is recorded and one complete measurement cycle is called a breaking/making trace. In the conductance-distance measurements, typical moving rate of piezo is 3.2 µm/s, which translates into an opening and corresponding closing rate for two leads of about 2-4 nm/s (attenuation factor ~ 0.001)[95]. The set up is capable of measuring a current ~ 1 nA at an applied bias of 1 V. Prior to the experiments, ~ 0.1 mM dichloromethane (DCM) solution of target molecules is deposited over the suspended part of the unbroken electrodes by means of self-assembly from the solution[81].

## 4.6. Theoretical calculation details
### 4.6.1. Ferrocene

Transport calculations are based on the DFT electronic structure of the ferrocene bound to a pair of Au clusters and NEGF formalism. These are implemented in TURBOMOLE[96] package and FHI-AIMS[97] package, with NEGF method in aitranss module of FHI-AIMS[90]. In both cases, we employed a PBE functional[98] with dispersion corrections (disp4[99] in TURBOMOLE or Van der Waals[100] and many body dispersion[101] in FHI-AIMS). Two different molecular geometries are considered – perpendicular and parallel geometries, as shown in figure 5. For each geometry, the optimal molecular position for several distances of Au electrodes is considered (see figure S9 in Supplementary Information). Furthermore, dependence on random ad-atoms that break the pyramidal symmetry is investigated and shown to be negligible, see Supplementary Information, Section (h) for details. The basis used is TZVP in TURBOMOLE and tight basis in FHI-AIMS.

The transmission function is also calculated for smaller basis set size, showing that the basis set size is converged – details are given in the Supplementary Information.

### 4.6.2. 1,1′-bis(aminomethyl)ferrocene and 1,1′-dicyanoferrocene

First-principles based density functional theory (DFT) calculations have been used for electronic structure calculations using Quantum ATK[102]. This study uses the spin-polarized generalized gradient approximation (SGGA) of the Perdew - Burke - Ernzerhof (PBE) exchange-correlation functional with double-zeta polarized basis set and a 120 Hartree density mesh cut-off. Geometry optimization for each structure has been performed for forces smaller than 10 meV/Å. The transmission probabilities T(E) of the 2-probe geometry was investigated using the Non-equilibrium Green's Function (NEGF) formalism in conjunction with DFT theory. For the transport calculation, a k-point grid of $3 \times 3 \times 134$ was used with 134 along the transport direction, as no changes in the transmission was detected beyond that limit.

**Acknowledgements:**

B. Pabi acknowledges the support from DST-Inspire fellowship, Government of India (Inspire code IF170934) and A. N. Pal acknowledges the funding from the Department of Science and Technology, Government of India (Grant No. CRG/2020/004208). The authors acknowledge clean room fabrication facilities of SNBNCBS, and Priya Mahadevan and Debayan Mondal for fruitful discussions. S. Marek acknowledges the support from the Ministry of Education, Youth and Sports of the Czech Republic though the e-INFRA CZ (ID:90140), which provided the computational resources. S. Marek also acknowledges financial support provided by the Charles University through GAUK (ID366222).

**Conflicts of interest:**

There are no conflicts of interest.

**References:**

(1) Aviram, A.; Ratner, M. A. Molecular Rectifiers. *Chem Phys Lett* **1974**, *29* (2), 277–283. https://doi.org/10.1016/0009-2614(74)85031-1.

(2) Reichert, J.; Ochs, R.; Beckmann, D.; Weber, H. B.; Mayor, M.; Löhneysen, H. v. Driving Current through Single Organic Molecules. *Phys Rev Lett* **2002**, *88* (17), 4. https://doi.org/10.1103/PhysRevLett.88.176804.

(3) Seferos, D. S.; Blum, A. S.; Kushmerick, J. G.; Bazan, G. C. Single-Molecule Charge-Transport Measurements That Reveal Technique-Dependent Perturbations. *J Am Chem Soc* **2006**, *128* (34), 11260–11267. https://doi.org/10.1021/ja062898j.

(4) Kergueris, C.; Bourgoin, J. P.; Esteve, D.; Urbina, C.; Magoga, M.; Joachim, C. Electron Transport through a Metal-Molecule-Metal Junction. *Phys Rev B - Condens Matter Mater Phys* **1999**, *59* (19), 12505–12513. https://doi.org/10.1103/PhysRevB.59.12505.

(5) Donhauser, Z. J.; Mantooth, B. A.; Kelly, K. F.; Bumm, L. A.; Monnell, J. D.; Stapleton, J. J.; Price, J.; Rawlett, A. M.; Allara, D. L.; Tour, J. M.; Weiss, P. S. Conductance Switching in Single Molecules through Conformational Changes. *Science (80- )* **2001**, *292* (5525), 2303–2307. https://doi.org/10.1126/science.1060294.

(6) Venkataraman, L.; Klare, J. E.; Nuckolls, C.; Hybertsen, M. S.; Steigerwald, M. L. Dependence of Single-Molecule Junction Conductance on Molecular Conformation. *Nature* **2006**, *442* (7105), 904–907. https://doi.org/10.1038/nature05037.

(7) Cui, X. D.; Primak, A.; Zarate, X.; Tomfohr, J.; Sankey, O. F.; Moore, A. L.; Moore, T. A.; Gust, D.; Harris, G.; Lindsay, S. M. Reproducible Measurement of Single-Molecule Conductivity. *Science (80- )* **2001**, *294* (5542), 571–574. https://doi.org/10.1126/science.1064354.

(8) Dadosh, T.; Gordin, Y.; Krahne, R.; Khivrich, I.; Mahalu, D.; Frydman, V.; Sperling, J.; Yacoby, A.; Bar-Joseph, I. Measurement of the Conductance of Single Conjugated Molecules. *Nature* **2005**, *436* (7051), 677–680. https://doi.org/10.1038/nature03898.

(9) Park, J.; Pasupathy, A. N.; Goldsmith, J. I.; Chang, C.; Yalsh, Y.; Petta, J. R.; Rinkoski, M.; Sethna, J. P.; Abruña, H. D.; McEuen, P. L.; Ralph, D. C. Coulomb Blockade and the Kondo Effect in Single-Atom Transistors. *Nature* **2002**, *417* (6890), 722–725. https://doi.org/10.1038/nature00791.

(10) Reed, M. A.; Zhou, C.; Muller, C. J.; Burgin, T. P.; Tour, J. M. Conductance of a Molecular Junction. *Science (80- )* **1997**, *278* (October), 1–3.

(11) Nikiforov, M. Single-Molecule Resistance Measured by Repeated Formation of Molecular Junctions. *MRS Bull* **2003**, *28* (11), 790–792. https://doi.org/10.1557/mrs2003.223.

(12) Fan, F. R. F.; Yang, J.; Cai, L.; Price, D. W.; Dirk, S. M.; Kosynkin, D. V.; Yao, Y.; Rawlett, A. M.; Tour, J. M.; Bard, A. J. Charge Transport through Self-Assembled Monolayers of Compounds of Interest in Molecular Electronics. *J Am Chem Soc* **2002**, *124* (19), 5550–5560. https://doi.org/10.1021/ja017706t.

(13) Andres, R. P.; Bein, T.; Dorogi, M.; Feng, S.; Henderson, J. I.; Kubiak, C. P.; Mahoney, W.; Osifchin, R. G.; Reifenberger, R. "Coulomb Staircase" at Room Temperature in a Self-Assembled Molecular Nanostructure. *Science (80- )* **1996**, *272* (5266), 1323–1325. https://doi.org/10.1126/science.272.5266.1323.

(14) Slowinski, K.; Fong, H. K. Y.; Majda, M. Mercury-Mercury Tunneling Junctions. 1. Electron Tunneling across Symmetric and Asymmetric Alkanethiolate Bilayers. *J Am Chem Soc* **1999**, *121* (31), 7257–7261. https://doi.org/10.1021/ja991613i.

(15) Li, C.; Pobelov, I.; Wandlowski, T.; Bagrets, A.; Arnold, A.; Evers, F. Charge Transport in Single Au | Alkanedithiol | Au Junctions: Coordination Geometries and Conformational Degrees of Freedom. *J Am Chem Soc* **2008**, *130* (1), 318–326. https://doi.org/10.1021/ja0762386.

(16) Osorio, E. A.; Bjørnholm, T.; Lehn, J. M.; Ruben, M.; Van Der Zant, H. S. J. Single-Molecule Transport in Three-Terminal Devices. *J Phys Condens Matter* **2008**, *20* (37).

https://doi.org/10.1088/0953-8984/20/37/374121.

(17) Chen, J.; Reed, M. A.; Rawlett, A. M.; Tour, J. M. Large On-off Ratios and Negative Differential Resistance in a Molecular Electronic Device. *Science (80- )* **1999**, *286* (5444), 1550–1552. https://doi.org/10.1126/science.286.5444.1550.

(18) Haag, R.; Rampi, M. A.; Holmlin, R. E.; Whitesides, G. M. Electrical Breakdown of Aliphatic and Aromatic Self-Assembled Monolayers Used as Nanometer-Thick Organic Dielectrics. *J Am Chem Soc* **1999**, *121* (34), 7895–7906. https://doi.org/10.1021/ja990230h.

(19) Smit, R. H. M.; Noat, Y.; Untiedt, C.; Lang, N. D.; Van Hemert, M. C.; Van Ruitenbeek, J. M. Measurement of the Conductance of a Hydrogen Molecule. *Nature* **2002**, *419* (6910), 906–909. https://doi.org/10.1038/nature01103.

(20) Zhou, P.; Zheng, J.; Han, T.; Chen, L.; Cao, W.; Zhu, Y.; Zhou, D.; Li, R.; Tian, Y.; Liu, Z.; Liu, J.; Hong, W. Electrostatic Gating of Single-Molecule Junctions Based on the STM-BJ Technique. *Nanoscale* **2021**, *13* (16), 7600–7605. https://doi.org/10.1039/d1nr00157d.

(21) Lo, E. Transport Properties of a Single-Molecule Diode. **2012**, No. 6, 4931–4939.

(22) Greenwald, J. E.; Cameron, J.; Findlay, N. J.; Fu, T.; Gunasekaran, S.; Skabara, P. J.; Venkataraman, L. Highly Nonlinear Transport across Single-Molecule Junctions via Destructive Quantum Interference. *Nat Nanotechnol* **2021**, *16* (3), 313–317. https://doi.org/10.1038/s41565-020-00807-x.

(23) Zang, Y.; Ray, S.; Fung, E. D.; Borges, A.; Garner, M. H.; Steigerwald, M. L.; Solomon, G. C.; Patil, S.; Venkataraman, L. Resonant Transport in Single Diketopyrrolopyrrole Junctions. *J Am Chem Soc* **2018**, *140* (41), 13167–13170. https://doi.org/10.1021/jacs.8b06964.

(24) Yelin, T.; Chakrabarti, S.; Vilan, A.; Tal, O. Richness of Molecular Junction Configurations Revealed by Tracking a Full Pull-Push Cycle. *Nanoscale* **2021**, *13* (44), 18434–18440. https://doi.org/10.1039/d1nr05680h.

(25) Luka-Guth, K.; Hambsch, S.; Bloch, A.; Ehrenreich, P.; Briechle, B. M.; Kilibarda, F.; Sendler, T.; Sysoiev, D.; Huhn, T.; Erbe, A.; Scheer, E. Role of Solvents in the Electronic Transport Properties of Single-Molecule Junctions. *Beilstein J Nanotechnol* **2016**, *7* (1), 1055–1067. https://doi.org/10.3762/bjnano.7.99.

(26) Bai, J.; Li, X.; Zhu, Z.; Zheng, Y.; Hong, W. Single-Molecule Electrochemical Transistors. *Adv Mater* **2021**, *33* (50). https://doi.org/10.1002/adma.202005883.

(27) Choi, B.; Capozzi, B.; Ahn, S.; Turkiewicz, A.; Lovat, G.; Nuckolls, C.; Steigerwald, M. L.; Venkataraman, L.; Roy, X. Solvent-Dependent Conductance Decay Constants in Single Cluster Junctions. *Chem Sci* **2016**, *7* (4), 2701–2705. https://doi.org/10.1039/c5sc02595h.

(28) Hu, Y.; Li, J.; Zhou, Y.; Shi, J.; Li, G.; Song, H.; Yang, Y.; Shi, J.; Hong, W. Single Dynamic Covalent Bond Tailored Responsive Molecular Junctions. *Angew Chemie - Int Ed*

**2021**, *60* (38), 20872–20878. https://doi.org/10.1002/anie.202106666.

(29) Zang, Y.; Zou, Q.; Fu, T.; Ng, F.; Fowler, B.; Yang, J.; Li, H.; Steigerwald, M. L.; Nuckolls, C.; Venkataraman, L. Directing Isomerization Reactions of Cumulenes with Electric Fields. *Nat Commun* **2019**, *10* (1), 1–7. https://doi.org/10.1038/s41467-019-12487-w.

(30) Schosser, W. M.; Hsu, C.; Zwick, P.; Beltako, K.; Dulic, D.; Mayor, M.; Van Der Zant, H. S. J.; Pauly, F. Mechanical Conductance Tunability of a Porphyrin-Cyclophane Single-Molecule Junction. *Nanoscale* **2022**, *14* (3), 984–992. https://doi.org/10.1039/d1nr06484c.

(31) Stefani, D.; Weiland, K. J.; Skripnik, M.; Hsu, C.; Perrin, M. L.; Mayor, M.; Pauly, F.; Van Der Zant, H. S. J. Large Conductance Variations in a Mechanosensitive Single-Molecule Junction. *Nano Lett* **2018**, *18* (9), 5981–5988. https://doi.org/10.1021/acs.nanolett.8b02810.

(32) Caneva, S.; Hermans, M.; Lee, M.; García-Fuente, A.; Watanabe, K.; Taniguchi, T.; Dekker, C.; Ferrer, J.; Van Der Zant, H. S. J.; Gehring, P. A Mechanically Tunable Quantum Dot in a Graphene Break Junction. *Nano Lett* **2020**, *20* (7), 4924–4931. https://doi.org/10.1021/acs.nanolett.0c00984.

(33) Hayakawa, R.; Karimi, M. A.; Wolf, J.; Huhn, T.; Zöllner, M. S.; Herrmann, C.; Scheer, E. Large Magnetoresistance in Single-Radical Molecular Junctions. *Nano Lett* **2016**, *16* (8), 4960–4967. https://doi.org/10.1021/acs.nanolett.6b01595.

(34) Chen, H.; Sangtarash, S.; Li, G.; Gantenbein, M.; Cao, W.; Alqorashi, A.; Liu, J.; Zhang, C.; Zhang, Y.; Chen, L.; Chen, Y.; Olsen, G.; Sadeghi, H.; Bryce, M. R.; Lambert, C. J.; Hong, W. Exploring the Thermoelectric Properties of Oligo(Phenylene-Ethynylene) Derivatives. *Nanoscale* **2020**, *12* (28), 15150–15156. https://doi.org/10.1039/d0nr03303k.

(35) Cui, L.; Jeong, W.; Hur, S.; Matt, M.; Klöckner, J. C.; Pauly, F.; Nielaba, P.; Cuevas, J. C.; Meyhofer, E.; Reddy, P. Quantized Thermal Transport in Single-Atom Junctions. *Science (80- )* **2017**, *355* (6330), 1192–1195. https://doi.org/10.1126/science.aam6622.

(36) Cui, L.; Hur, S.; Akbar, Z. A.; Klöckner, J. C.; Jeong, W.; Pauly, F.; Jang, S. Y.; Reddy, P.; Meyhofer, E. Thermal Conductance of Single-Molecule Junctions. *Nature* **2019**, *572* (7771), 628–633. https://doi.org/10.1038/s41586-019-1420-z.

(37) Reddy, P.; Jang, S. Y.; Segalman, R. A.; Majumdar, A. Thermoelectricity in Molecular Junctions. *Science (80- )* **2007**, *315* (5818), 1568–1571. https://doi.org/10.1126/science.1137149.

(38) Kim, T.; Darancet, P.; Widawsky, J. R.; Kotiuga, M.; Quek, S. Y.; Neaton, J. B.; Venkataraman, L. Determination of Energy Level Alignment and Coupling Strength in 4,4′-Bipyridine Single-Molecule Junctions. *Nano Lett* **2014**, *14* (2), 794–798. https://doi.org/10.1021/nl404143v.

(39) Paoletta, A. L.; Fung, E. D.; Venkataraman, L. Gap Size-Dependent Plasmonic Enhancement in Electroluminescent Tunnel Junctions. *ACS Photonics* **2022**, *9* (2), 688–693. https://doi.org/10.1021/acsphotonics.1c01757.

(40) Marqués-González, S.; Matsushita, R.; Kiguchi, M. Surface Enhanced Raman Scattering of Molecules in Metallic Nanogaps. *J Opt (United Kingdom)* **2015**, *17* (11). https://doi.org/10.1088/2040-8978/17/11/114001.

(41) Reed, M. A.; Zhou, C.; Muller, C. J.; Burgin, T. P.; Tour, J. M. Conductance of a Molecular Junction. *Science (80- )* **1997**, *278* (5336), 252–254. https://doi.org/10.1126/science.278.5336.252.

(42) Ulrich, J.; Esrail, D.; Pontius, W.; Venkataraman, L.; Millar, D.; Doerrer, L. H. Variability of Conductance in Molecular Junctions. *J Phys Chem B* **2006**, *110* (6), 2462–2466. https://doi.org/10.1021/jp056455y.

(43) Quek, S. Y.; Venkataraman, L.; Choi, H. J.; Louie, S. G.; Hybertsen, M. S.; Neaton, J. B. Amine - Gold Linked Single-Molecule Circuits: Experiment and Theory. *Nano Lett* **2007**, *7* (11), 3477–3482. https://doi.org/10.1021/nl072058i.

(44) Venkataraman, L.; Klare, J. E.; Tam, I. W.; Nuckolls, C.; Hybertsen, M. S.; Steigerwald, M. L. Single-Molecule Circuits with Well-Defined Molecular Conductance. *Nano Lett* **2006**, *6* (3), 458–462. https://doi.org/10.1021/nl052373+.

(45) Mishchenko, A.; Zotti, L. A.; Vonlanthen, D.; Bürkle, M.; Pauly, F.; Cuevas, J. C.; Mayor, M.; Wandlowski, T. Single-Molecule Junctions Based on Nitrile-Terminated Biphenyls: A Promising New Anchoring Group. *J Am Chem Soc* **2011**, *133* (2), 184–187. https://doi.org/10.1021/ja107340t.

(46) Lörtscher, E.; Cho, C. J.; Mayor, M.; Tschudy, M.; Rettner, C.; Riel, H. Influence of the Anchor Group on Charge Transport through Single-Molecule Junctions. *ChemPhysChem* **2011**, *12* (9), 1677–1682. https://doi.org/10.1002/cphc.201000960.

(47) Kim, B.; Beebe, J. M.; Jun, Y.; Zhu, X. Y.; Frisbie, G. D. Correlation between HOMO Alignment and Contact Resistance in Molecular Junctions: Aromatic Thiols versus Aromatic Isocyanides. *J Am Chem Soc* **2006**, *128* (15), 4970–4971. https://doi.org/10.1021/ja0607990.

(48) Chen, F.; Li, X.; Hihath, J.; Huang, Z.; Tao, N. Effect of Anchoring Groups on Single-Molecule Conductance: Comparative Study of Thiol-, Amine-, and Carboxylic-Acid-Terminated Molecules. *J Am Chem Soc* **2006**, *128* (49), 15874–15881. https://doi.org/10.1021/ja065864k.

(49) Zotti, L. A.; Kirchner, T.; Cuevas, J. C.; Pauly, F.; Huhn, T.; Scheer, E.; Erbe, A. Revealing the Role of Anchoring Groups in the Electrical Conduction through Single-Molecule Junctions. *Small* **2010**, *6* (14), 1529–1535. https://doi.org/10.1002/smll.200902227.

(50) Cheng, Z. L.; Skouta, R.; Vazquez, H.; Widawsky, J. R.; Schneebeli, S.; Chen, W.; Hybertsen, M. S.; Breslow, R.; Venkataraman, L. In Situ Formation of Highly Conducting Covalent Au-C Contacts for Single-Molecule Junctions. *Nat Nanotechnol* **2011**, *6* (6), 353–357. https://doi.org/10.1038/nnano.2011.66.

(51) Schull, G.; Frederiksen, T.; Arnau, A.; Sánchez-Portal, D.; Berndt, R. Atomic-Scale Engineering of Electrodes for Single-Molecule Contacts. *Nat Nanotechnol* **2011**, *6* (1), 23–27. https://doi.org/10.1038/nnano.2010.215.

(52) Martin, C. A.; Ding, D.; Sørensen, J. K.; Bjørnholm, T.; Van Ruitenbeek, J. M.; Van Der Zant, H. S. J. Fullerene-Based Anchoring Groups for Molecular Electronics. *J Am Chem Soc* **2008**, *130* (40), 13198–13199. https://doi.org/10.1021/ja804699a.

(53) Tada, T.; Yoshizawa, K. Quantum Transport Effects in Nanosized Graphite Sheets. II. Enhanced Transport Effects by Heteroatoms. *J Phys Chem B* **2003**, *107* (34), 8789–8793. https://doi.org/10.1021/jp021739t.

(54) Kiguchi, M.; Murakoshi, K. Highly Conductive Single Molecular Junctions by Direct Binding of π-Conjugated Molecule to Metal Electrodes. *Thin Solid Films* **2009**, *518* (2), 466–469. https://doi.org/10.1016/j.tsf.2009.07.024.

(55) Li, L.; Low, J. Z.; Wilhelm, J.; Liao, G.; Gunasekaran, S.; Prindle, C. R.; Starr, R. L.; Golze, D.; Nuckolls, C.; Steigerwald, M. L.; Evers, F.; Campos, L. M.; Yin, X.; Venkataraman, L. Insulators Based on Mono- and Di-Radical Cations. *Nat Chem* **2022**. https://doi.org/10.1038/s41557-022-00978-1.

(56) Yelin, T.; Korytár, R.; Sukenik, N.; Vardimon, R.; Kumar, B.; Nuckolls, C.; Evers, F.; Tal, O. Conductance Saturation in a Series of Highly Transmitting Molecular Junctions. *Nat Mater* **2016**, *15* (4), 444–449. https://doi.org/10.1038/nmat4552.

(57) Kaneko, S.; Nakazumi, T.; Kiguchi, M. Fabrication of a Well-Defined Single Benzene Molecule Junction Using Ag Electrodes. *J Phys Chem Lett* **2010**, *1* (24), 3520–3523. https://doi.org/10.1021/jz101506u.

(58) Schneebeli, S. T.; Kamenetska, M.; Cheng, Z.; Skouta, R.; Friesner, R. A.; Venkataraman, L.; Breslow, R. Single-Molecule Conductance through Multiple π-π-Stacked Benzene Rings Determined with Direct Electrode-to-Benzene Ring Connections. *J Am Chem Soc* **2011**, *133* (7), 2136–2139. https://doi.org/10.1021/ja111320n.

(59) Kiguchi, M.; Tal, O.; Wohlthat, S.; Pauly, F.; Krieger, M.; Djukic, D.; Cuevas, J. C.; Van Ruitenbeek, J. M. Highly Conductive Molecular Junctions Based on Direct Binding of Benzene to Platinum Electrodes. *Phys Rev Lett* **2008**, *101* (4), 1–4. https://doi.org/10.1103/PhysRevLett.101.046801.

(60) Tal, O.; Kiguchi, M.; Thijssen, W. H. A.; Djukic, D.; Untiedt, C.; Smit, R. H. M.; Van Ruitenbeek, J. M. Molecular Signature of Highly Conductive Metal-Molecule-Metal Junctions. *Phys Rev B - Condens Matter Mater Phys* **2009**, *80* (8), 1–8. https://doi.org/10.1103/PhysRevB.80.085427.

(61) Diez-Perez, I.; Hihath, J.; Hines, T.; Wang, Z. S.; Zhou, G.; Müllen, K.; Tao, N. Controlling Single-Molecule Conductance through Lateral Coupling of φ Orbitals. *Nat Nanotechnol* **2011**, *6* (4), 226–231. https://doi.org/10.1038/nnano.2011.20.

(62) Kuang, G.; Chen, S. Z.; Wang, W.; Lin, T.; Chen, K.; Shang, X.; Liu, P. N.; Lin, N. Resonant Charge Transport in Conjugated Molecular Wires beyond 10 Nm Range. *J Am Chem Soc* **2016**, *138* (35), 11140–11143. https://doi.org/10.1021/jacs.6b07416.

(63) Li, S.; Yu, H.; Li, J.; Angello, N.; Jira, E. R.; Li, B.; Burke, M. D.; Moore, J. S.; Schroeder, C. M. Transition between Nonresonant and Resonant Charge Transport in Molecular Junctions. *Nano Lett* **2021**, *21* (19), 8340–8347. https://doi.org/10.1021/acs.nanolett.1c02915.

(64) Song, H.; Kim, Y.; Jang, Y. H.; Jeong, H.; Reed, M. A.; Lee, T. Observation of Molecular Orbital Gating. *Nature* **2009**, *462* (7276), 1039–1043. https://doi.org/10.1038/nature08639.

(65) Zang, Y.; Fung, E. D.; Fu, T.; Ray, S.; Garner, M. H.; Borges, A.; Steigerwald, M. L.; Patil, S.; Solomon, G.; Venkataraman, L. Voltage-Induced Single-Molecule Junction Planarization. *Nano Lett* **2021**, *21* (1), 673–679. https://doi.org/10.1021/acs.nanolett.0c04260.

(66) Zang, Y.; Pinkard, A.; Liu, Z. F.; Neaton, J. B.; Steigerwald, M. L.; Roy, X.; Venkataraman, L. Electronically Transparent Au-N Bonds for Molecular Junctions. *J Am Chem Soc* **2017**, *139* (42), 14845–14848. https://doi.org/10.1021/jacs.7b08370.

(67) Bredow, T.; Tegenkamp, C.; Pfnür, H.; Meyer, J.; Maslyuk, V. V.; Mertig, I. Ferrocene-1,1′-Dithiol as Molecular Wire between Ag Electrodes: The Role of Surface Defects. *J Chem Phys* **2008**, *128* (6). https://doi.org/10.1063/1.2827867.

(68) Zhou, L.; Yang, S. W.; Ng, M. F.; Sullivan, M. B.; Tan, V. B. C.; Shen, L. One-Dimensional Iron-Cyclopentadienyl Sandwich Molecular Wire with Half Metallic, Negative Differential Resistance and High-Spin Filter Efficiency Properties. *J Am Chem Soc* **2008**, *130* (12), 4023–4027. https://doi.org/10.1021/ja7100246.

(69) Zhao, X.; Stadler, R. DFT-Based Study of Electron Transport through Ferrocene Compounds with Different Anchor Groups in Different Adsorption Configurations of an STM Setup. *Phys Rev B* **2019**, *99* (4), 1–10. https://doi.org/10.1103/PhysRevB.99.045431.

(70) Zhao, X.; Kastlunger, G.; Stadler, R. Quantum Interference in Coherent Tunneling through Branched Molecular Junctions Containing Ferrocene Centers. *Phys Rev B* **2017**, *96* (8), 1–12. https://doi.org/10.1103/PhysRevB.96.085421.

(71) Kanthasamy, K.; Ring, M.; Nettelroth, D.; Tegenkamp, C.; Butenschön, H.; Pauly, F.; Pfnür, H. Charge Transport through Ferrocene 1,1′-Diamine Single-Molecule Junctions. *Small* **2016**, *12* (35), 4849–4856. https://doi.org/10.1002/smll.201601051.

(72) Cristian Morari, Ivan Rungger, Alexandre R. Rocha, Stefano Sanvito, Sorin Melinte, A.; Rignanese, G.-M. Electronic Transport Properties of 1,1=-Ferrocene Dicarboxylic Acid Linked to Al(111) Electrodes. *ACS Nano* **209AD**, *3* (12), 4147–4143. https://doi.org/10.1016/j.comptc.2015.10.025.

(73) Aragonès, A. C.; Darwish, N.; Ciampi, S.; Jiang, L.; Roesch, R.; Ruiz, E.; Nijhuis, C. A.;

Díez-Pérez, I. Control over Near-Ballistic Electron Transport through Formation of Parallel Pathways in a Single-Molecule Wire. *J Am Chem Soc* **2019**, *141* (1), 240–250. https://doi.org/10.1021/jacs.8b09086.

(74) Pal, A. N.; Li, D.; Sarkar, S.; Chakrabarti, S.; Vilan, A.; Kronik, L.; Smogunov, A.; Tal, O. Nonmagnetic Single-Molecule Spin-Filter Based on Quantum Interference. *Nat Commun* **2019**, *10* (1), 5565. https://doi.org/10.1038/s41467-019-13537-z.

(75) Getty, S. A.; Engtrakul, C.; Wang, L.; Liu, R.; Ke, S. H.; Baranger, H. U.; Yang, W.; Fuhrer, M. S.; Sita, L. R. Near-Perfect Conduction through a Ferrocene-Based Molecular Wire. *Phys Rev B - Condens Matter Mater Phys* **2005**, *71* (24), 2–5. https://doi.org/10.1103/PhysRevB.71.241401.

(76) Lawson, B.; Zahl, P.; Hybertsen, M. S.; Kamenetska, M. Formation and Evolution of Metallocene Single-Molecule Circuits with Direct Gold-π Links. *J Am Chem Soc* **2022**, *144* (14), 6504–6515. https://doi.org/10.1021/jacs.2c01322.

(77) Liu, R.; Han, Y.; Sun, F.; Khatri, G.; Kwon, J.; Nickle, C.; Wang, L.; Thompson, D.; Li, Z.; Nijhuis, C. A.; Barco, E.; Liu, R.; Khatri, G.; Kwon, J.; Nickle, C.; Barco, P. E. Stable Universal 1- and 2-Input Single-Molecule Logic Gates. https://doi.org/10.1002/adma.202202135.

(78) Trouwborst, M. L.; Huisman, E. H.; Bakker, F. L.; Van Der Molen, S. J.; Van Wees, B. J. Single Atom Adhesion in Optimized Gold Nanojunctions. *Phys Rev Lett* **2008**, *100* (17). https://doi.org/10.1103/PhysRevLett.100.175502.

(79) Pabi, B.; Mondal, D.; Mahadevan, P.; Pal, A. N. Probing Metal-Molecule Contact at the Atomic Scale via Conductance Jumps. *Phys Rev B* **2021**, *104* (12), 1–8. https://doi.org/10.1103/PhysRevB.104.L121407.

(80) Agraït, N.; Yeyati, A. L.; van Ruitenbeek, J. M. Quantum Properties of Atomic-Sized Conductors. *Phys Rep* **2003**, *377* (2–3), 81–279. https://doi.org/10.1016/S0370-1573(02)00633-6.

(81) Perrin, M. L.; Martin, C. A.; Prins, F.; Shaikh, A. J.; Eelkema, R.; van Esch, J. H.; van Ruitenbeek, J. M.; van der Zant, H. S. J.; Dulić, D. Charge Transport in a Zinc-Porphyrin Single-Molecule Junction. *Beilstein J Nanotechnol* **2011**, *2* (1), 714–719. https://doi.org/10.3762/bjnano.2.77.

(82) Untiedt, C.; Yanson, A. I.; Grande, R.; Rubio-Bollinger, G.; Agraït, N.; Vieira, S.; van Ruitenbeek, J. M. Calibration of the Length of a Chain of Single Gold Atoms. *Phys Rev B - Condens Matter Mater Phys* **2002**, *66* (8), 854181–854186. https://doi.org/10.1103/PhysRevB.66.085418.

(83) Balogh, Z.; Visontai, D.; Makk, P.; Gillemot, K.; Oroszlány, L.; Pósa, L.; Lambert, C.; Halbritter, A. Precursor Configurations and Post-Rupture Evolution of Ag-CO-Ag Single-Molecule Junctions. *Nanoscale* **2014**, *6* (24), 14784–14791. https://doi.org/10.1039/c4nr04645e.

(84) Pal, A. N.; Klein, T.; Vilan, A.; Tal, O. Electronic Conduction during the Formation Stages of a Single-Molecule Junction. *Beilstein J Nanotechnol* **2018**, *9* (1), 1471–1477. https://doi.org/10.3762/bjnano.9.138.

(85) Balogh, Z.; Makk, P.; Halbritter, A. Alternative Types of Molecule-Decorated Atomic Chains in Au-CO-Au Single-Molecule Junctions. *Beilstein J Nanotechnol* **2015**, *6* (1), 1369–1376. https://doi.org/10.3762/bjnano.6.141.

(86) Makk, P.; Tomaszewski, D.; Martinek, J.; Balogh, Z.; Csonka, S.; Wawrzyniak, M.; Frei, M.; Venkataraman, L.; Halbritter, A. Correlation Analysis of Atomic and Single-Molecule Junction Conductance. *ACS Nano* **2012**, *6* (4), 3411–3423. https://doi.org/10.1021/nn300440f.

(87) Camarasa-Gómez, M.; Hernangómez-Pérez, D.; Inkpen, M. S.; Lovat, G.; Fung, E. D.; Roy, X.; Venkataraman, L.; Evers, F. Mechanically Tunable Quantum Interference in Ferrocene-Based Single-Molecule Junctions. *Nano Lett* **2020**, *20* (9), 6381–6386. https://doi.org/10.1021/acs.nanolett.0c01956.

(88) Hong, W.; Manrique, D. Z.; Moreno-García, P.; Gulcur, M.; Mishchenko, A.; Lambert, C. J.; Bryce, M. R.; Wandlowski, T. Single Molecular Conductance of Tolanes: Experimental and Theoretical Study on the Junction Evolution Dependent on the Anchoring Group. *J Am Chem Soc* **2012**, *134* (4), 2292–2304. https://doi.org/10.1021/ja209844r.

(89) Moreno-garc, P.; Gulcur, M.; Manrique, D. Z.; Pope, T.; Hong, W.; Kaliginedi, V.; Huang, C.; Batsanov, A. S.; Bryce, M. R.; Lambert, C.; Wandlowski, T. Single-Molecule Conductance of Functionalized Oligoynes: Length Dependence and Junction Evolution. **2013**.

(90) Arnold, A.; Weigend, F.; Evers, F. Quantum Chemistry Calculations for Molecules Coupled to Reservoirs: Formalism, Implementation, and Application to Benzenedithiol. *J Chem Phys* **2007**, *126* (17). https://doi.org/10.1063/1.2716664.

(91) Patti, A.; Pedotti, S. Synthesis of Hybrid Ferrocene-Proline Amides as Active Catalysts for Asymmetric Aldol Reactions in Water. *European J Org Chem* **2014**, *2014* (3), 624–630. https://doi.org/10.1002/ejoc.201301346.

(92) Strehler, F.; Hildebrandt, A.; Korb, M.; Rüffer, T.; Lang, H. From Ferrocenecarbonitriles to Ferrocenylimines: Synthesis, Structure, and Reaction Chemistry. *Organometallics* **2014**, *33* (16), 4279–4289. https://doi.org/10.1021/om500597c.

(93) C. J. Muller, J. M. van Ruitenbeek, and L. J. de J. Conductance and Supercurrent Discontinuities in Atomic-Scale Metallic Constrictions of Variable Width C. *Phys Rev Lett* **1992**, *69* (1), 140–143.

(94) Pabi, B.; Pal, A. N. An Experimental Set up to Probe the Quantum Transport through Single Atomic/Molecular Junction at Room Temperature. *Pramana* **2022**. https://doi.org/10.1007/s12043-022-02489-7.

(95) Vrouwe, S. A. G.; Van Der Giessen, E.; Van Der Molen, S. J.; Dulic, D.; Trouwborst, M. L.; Van Wees, B. J. Mechanics of Lithographically Defined Break Junctions. *Phys Rev B - Condens Matter Mater Phys* **2005**, *71* (3), 1–7. https://doi.org/10.1103/PhysRevB.71.035313.

(96) Balasubramani, S. G.; Chen, G. P.; Coriani, S.; Diedenhofen, M.; Frank, M. S.; Franzke, Y. J.; Furche, F.; Grotjahn, R.; Harding, M. E.; Hättig, C.; Hellweg, A.; Helmich-Paris, B.; Holzer, C.; Huniar, U.; Kaupp, M.; Marefat Khah, A.; Karbalaei Khani, S.; Müller, T.; Mack, F.; Nguyen, B. D.; Parker, S. M.; Perlt, E.; Rappoport, D.; Reiter, K.; Roy, S.; Rückert, M.; Schmitz, G.; Sierka, M.; Tapavicza, E.; Tew, D. P.; Van Wüllen, C.; Voora, V. K.; Weigend, F.; Wodyński, A.; Yu, J. M. TURBOMOLE: Modular Program Suite for Ab Initio Quantum-Chemical and Condensed-Matter Simulations. *J Chem Phys* **2020**, *152* (18). https://doi.org/10.1063/5.0004635.

(97) Blum, V.; Gehrke, R.; Hanke, F.; Havu, P.; Havu, V.; Ren, X.; Reuter, K.; Scheffler, M. Ab Initio Molecular Simulations with Numeric Atom-Centered Orbitals. *Comput Phys Commun* **2009**, *180* (11), 2175–2196. https://doi.org/10.1016/j.cpc.2009.06.022.

(98) Perdew, J. P.; Burke, K.; Ernzerhof, M. Generalized Gradient Approximation Made Simple. *Phys Rev Lett* **1996**, *77* (18), 3865–3868. https://doi.org/10.1103/PhysRevLett.77.3865.

(99) Caldeweyher, E.; Ehlert, S.; Hansen, A.; Neugebauer, H.; Spicher, S.; Bannwarth, C.; Grimme, S. A Generally Applicable Atomic-Charge Dependent London Dispersion Correction. *J Chem Phys* **2019**, *150* (15). https://doi.org/10.1063/1.5090222.

(100) Tkatchenko, A.; Scheffler, M. Accurate Molecular van Der Waals Interactions from Ground-State Electron Density and Free-Atom Reference Data. *Phys Rev Lett* **2009**, *102* (7), 6–9. https://doi.org/10.1103/PhysRevLett.102.073005.

(101) Tkatchenko, A.; Distasio, R. A.; Car, R.; Scheffler, M. Accurate and Efficient Method for Many-Body van Der Waals Interactions. *Phys Rev Lett* **2012**, *108* (23), 1–5. https://doi.org/10.1103/PhysRevLett.108.236402.

(102) Smidstrup, S.; Markussen, T.; Vancraeyveld, P.; Wellendorff, J.; Schneider, J.; Gunst, T.; Verstichel, B.; Stradi, D.; Khomyakov, P. A.; Vej-Hansen, U. G.; Lee, M. E.; Chill, S. T.; Rasmussen, F.; Penazzi, G.; Corsetti, F.; Ojanperä, A.; Jensen, K.; Palsgaard, M. L. N.; Martinez, U.; Blom, A.; Brandbyge, M.; Stokbro, K. QuantumATK: An Integrated Platform of Electronic and Atomic-Scale Modelling Tools. *J Phys Condens Matter* **2020**, *32* (1). https://doi.org/10.1088/1361-648X/ab4007.

# Supplementary Information

# Resonant transport in a highly conducting single molecular junction via metal-metal bond

Biswajit Pabi[1], Štepán Marek[2], Adwitiya Pal[3], Puja Kumari[4], Soumy Jyoti Ray[4], Arunabha Thakur[3], Richard Korytár[2], and Atindra Nath Pal[1*]

[1]*Department of Condensed Matter Physics and Material Sciences, S. N. Bose National Centre for Basic Sciences, Sector III, Block JD, Salt Lake, Kolkata 700106, India.*
[2]*Department of Condensed Matter Physics, Faculty of Mathematics and Physics, Charles University, 121 16, Prague 2, Czech Republic*
[3]*Department of Chemistry, Jadavpur University, Kolkata-700032, India.*
[4]*Department of Physics, Indian Institute of Technology Patna, Bihar- 801106, India.*

**Contents:-**
1. **Characterization of molecules**
    (a) 1,1′-bis(aminomethyl)ferrocene.
    (b) 1,1′-dicyanoferrocene.
2. **Analysis tools of experimental data**
    (a) Conductance histogram.
    (b) 2D conductance – distance histogram and average traces.
    (c) Stretching length histogram.
    (d) Correlation analysis.
3. **Additional experimental data**
    (a) Control experimental data.
    (b) Bias dependent conductance histogram of Au/ferrocene.
    (c) Precursor configuration.
    (d) Conditional analysis.
4. **Additional theoretical results**
    (a) Binding energy curve.
    (b) Molecular orbitals of free ferrocene.
    (c) Conductance of perpendicular and parallel geometry with applied voltage.
    (d) Effect of ring rotation of Cp rings on transmission.
    (e) Transmission function of perpendicular and parallel geometry.
    (f) Influence of Van der Waals correction and nature of adsorption.
    (g) Local Density of States on the Iron Ion
    (h) Stability of Observables – Basis Set Size and Ad-Atoms
    (i) Calculation of transmission function constrained to a single scattering state.
    (j) Coordinates of Converged Junctions
    (k) Binding energy of anchoring group coupled FC-$NH_2$ and FC-CN.

## 1. Characterization of molecules:
### (a) 1,1′-bis(aminomethyl)ferrocene

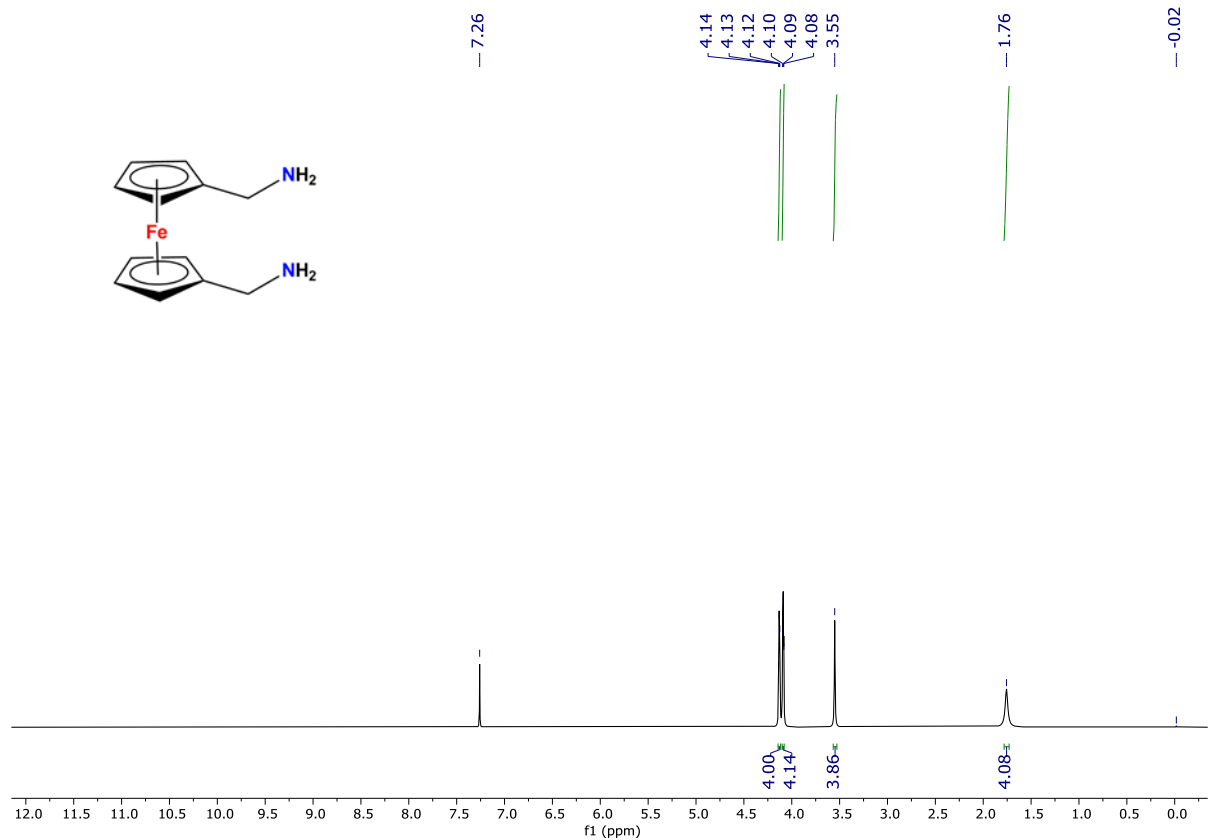

*Figure S1:* $^1H$ NMR spectrum of 1,1'-bis(aminomethyl)ferrocene in CDCl$_3$.

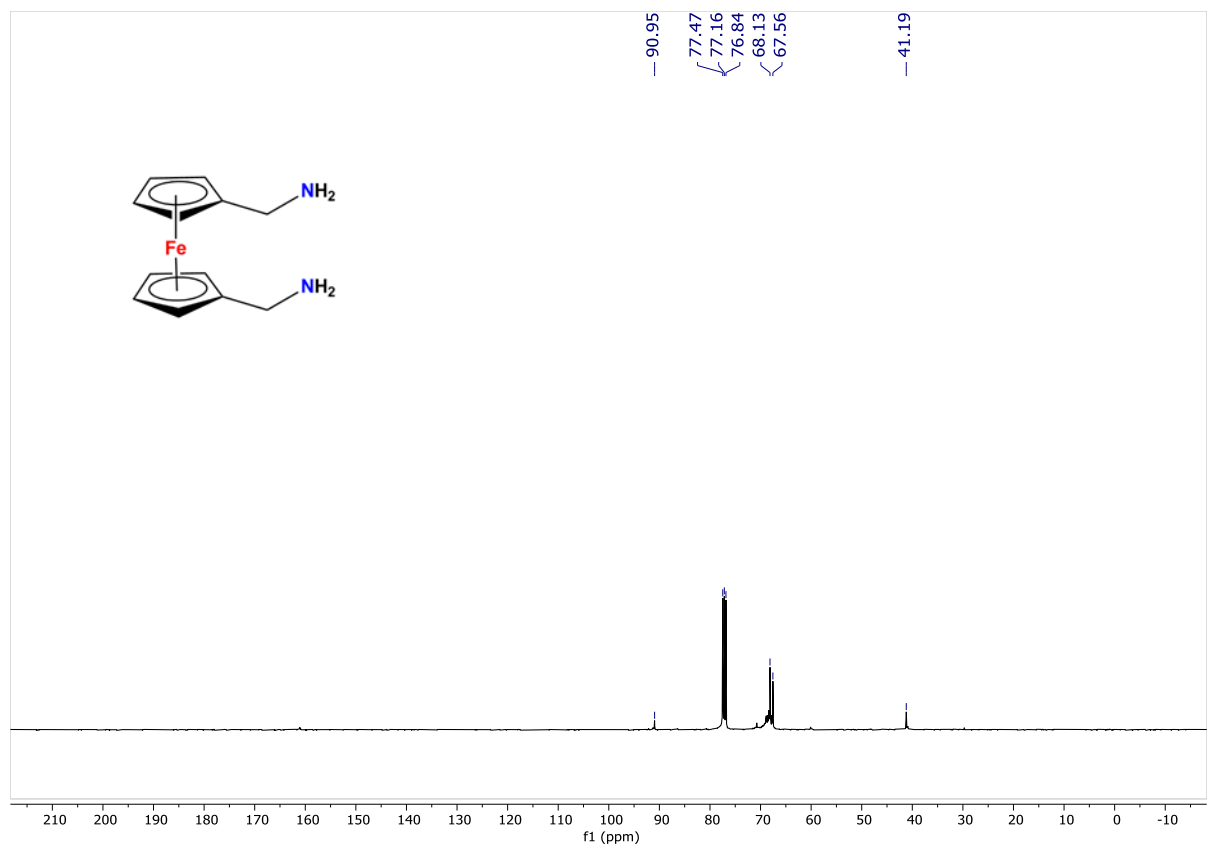

*Figure S2:* $^{13}$C NMR spectrum of 1,1'-bis(aminomethyl)ferrocene in CDCl$_3$.

**(b) 1,1'-dicyanoferrocene**

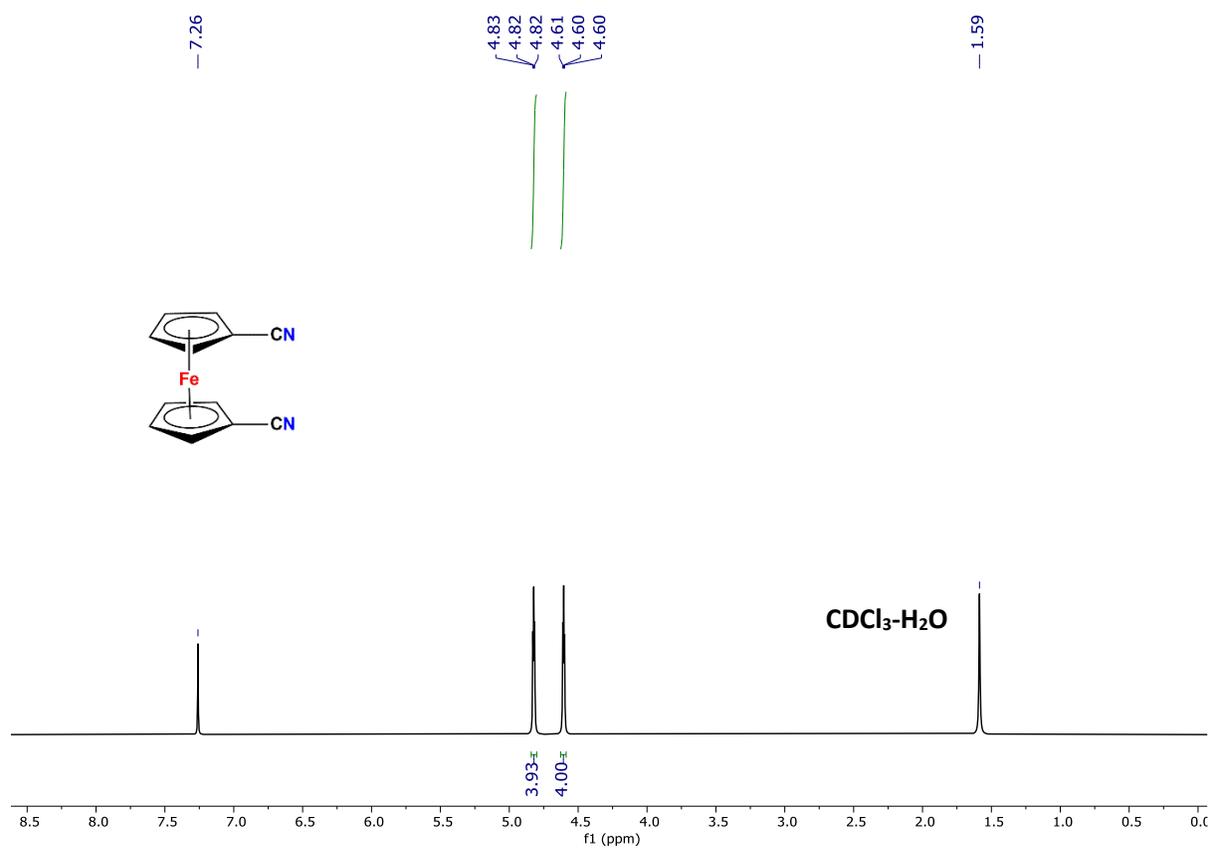

*Figure S3:* $^1$H NMR spectrum of 1,1'-dicyanoferrocene in CDCl$_3$.

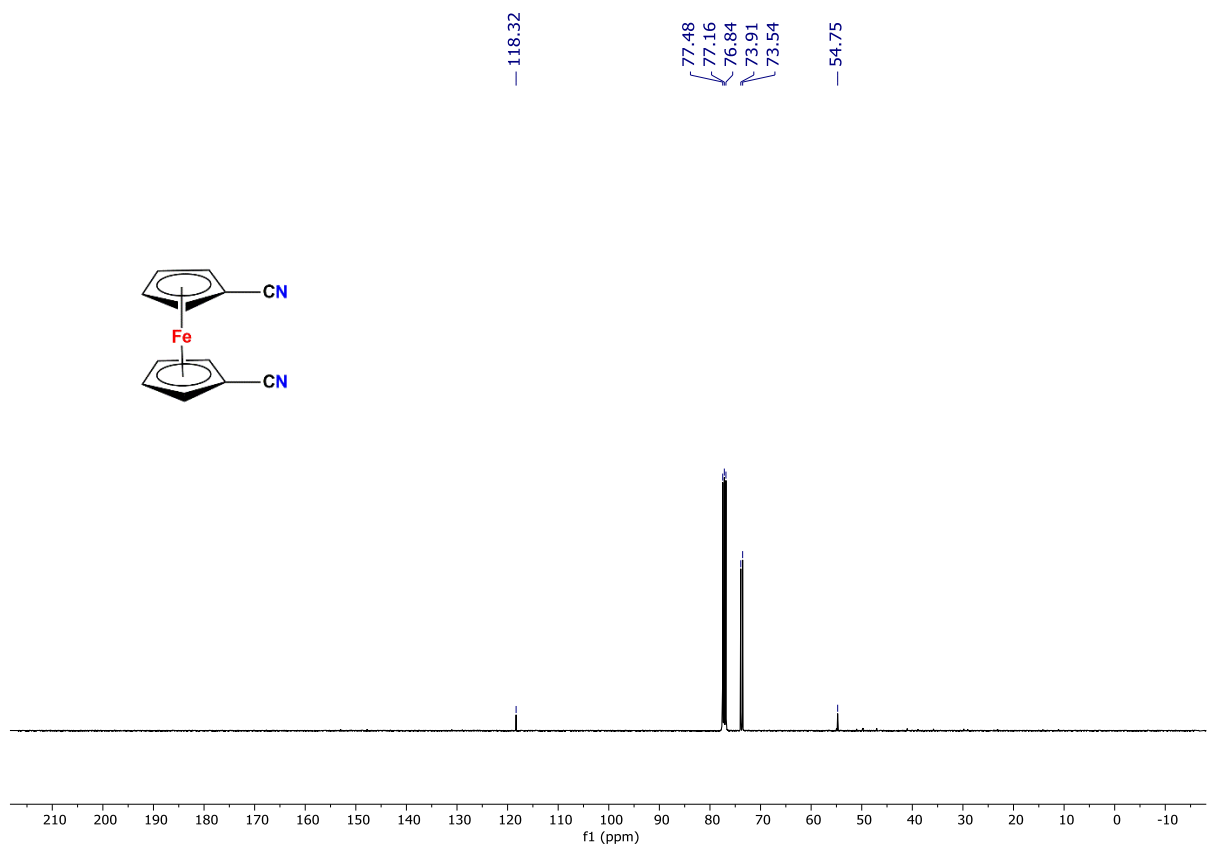

***Figure S4:*** *$^{13}$C NMR spectrum of 1,1'-dicyanoferrocene in CDCl$_3$.*

## 2. Analysis tools of experimental data:
### (a) 1D conductance histogram

In case of break junction experiment, conductance of the junction is measured as a function of electrode separation, starting from few tenths of atom to the breaking of the junction, by a small and constant increment of electrode separation. A complete measurement cycle is called a *conductance traces* and thousands or even more such traces are collected to look into the junction statistically. Before entering into the next cycle, electrodes are always crashed up to few atoms to delete the memory from previous cycle. In each cycle, junction evolves via series of elastic and plastic deformation leading to a unique configuration, which is beyond to probe experimentally. Thus, statistical analysis of large traces is critically important to identify the most possible scenario. This has been performed by creating conductance histogram[1] which presents distribution of conductance values collected from bulk number of traces. Consequently, when junction exhibits constant conductance values with respect to stretching (i.e. conductance plateaus), many values will add up to a small range of conductance and reflected as a peak in the histogram. Peaks in the histogram are fitted to calculate the most probable conductance values.

### (b) 2D conductance-distance histogram and average traces

As mentioned earlier, conductance of the junction is recorded as a function of electrode separation. However, conductance histogram provides only the conductance value, ignoring information related to the conductance in conjugation with electrode separation or distance. 2D conductance-distance histogram[2,3] is thus created to check the correlation of conductance and distance in a statistical manner. 2D conductance- distance histogram is constructed by following the way: first a conductance value is assigned for each traces as the origin of the distance axis i.e. zero distance point (for our case it is at 2 $G_0$). Each point in the traces is then contributes to one of the 2D bins of the histogram, defined by conductance and distance from the starting point. The resulting histogram can then be considered as a stack of many traces, placed on top of each other and is helpful to determine the most common feature during stretching.

Average traces for a conductance peak is prepared by calculating most probable conductance values in a certain conductance range, defined by the two local minima around the peak. This gives an additional information regarding the evolution of the junction. Linear fitting of the average traces yields the slope, useful to get an intuitive understanding about the metal-molecule coupling.

### (c) Stretching-Length histogram

Stretching length is defined as the difference between two absolute distance values, corresponding to a conductance $G_i$ (= 0.5 $G_0$; our case) to a desired conductance value $G_f$. In our case, $G_f$ is assigned at the one order of magnitude beneath the most probable conductance value of the corresponding conductance peak. Stretching length histogram is thus presents the distribution of stretching length, obtained from each traces. Most probable stretching length is calculated by Gaussian fitting of the histogram, which is an indicative of the stability of the junction[4]. Moreover, detailed analysis of stretching length histogram paves the way to quantify the junction formation probability[5].

### (d) Correlation analysis

Correlation analysis, introduced by the Makk and coworkers[6], is an extremely useful tool to detect the several features of junction formation and evolution, which cannot be accessible using conventional conductance histogram. To obtain the statistical relation between different junction configurations having its characteristics conductance value, correlation parameter can be defined as,

$$C_{m,n} = \frac{<\delta N_m(r) * \delta N_n(r)>_r}{\sqrt{<[\delta N_m(r)]^2>_r * <[\delta N_n(r)]^2>_r}}$$

Where $N_m(r)$ and $N_n(r)$ are the number of data points in the $m^{th}$ and $n^{th}$ bin of the given trace $r$. $\delta N_{m/n}(r) = N_{m/n}(r) - <N_{m/n}(r)>$ is the deviation from the mean value. Let's discuss the value of $C_{m,n}$ along with its physical significance-

(i) $C_{m,n}$ 0; Configurations are statistically independent.
(ii) $C_{m,n} \neq 0$; Configurations are statistically dependent and type of dependency is determined by the sign of the function. Positive values of $C_{m,n}$ leads to the positive correlation and for these cases configurations either appear or disappear together. Negative value of the function indicate negative correlation and more than average counts in one configuration is supported by the less than average counts in another configuration. It may also happen that formation of one configuration resists the formation of other.

Further details about the correlation analysis is described in reference 6. Using this method, a correlation map is drawn for arbitrary conductance pairs $G_m$ and $G_n$. Two axis corresponds to the two conductance values and color illustrates $C_{m,n}$.

## 3. Additional experimental data:
### (a) Control experimental data

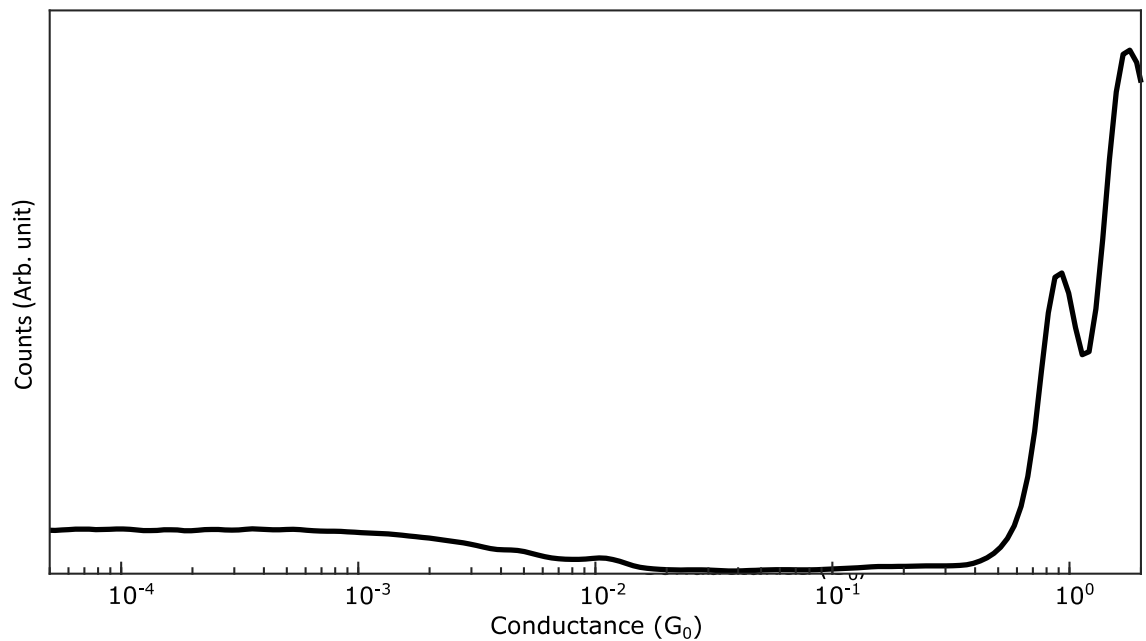

*Figure S5:* Conductance histogram constructed from 3000 consecutive conductance traces, recoded for the junction exposed to dichloromethane (DCM) solvent during breaking. Characteristics molecular signature is absent here, similar to reference 7.

### (b) Bias dependent conductance histogram of Au/ferrocene

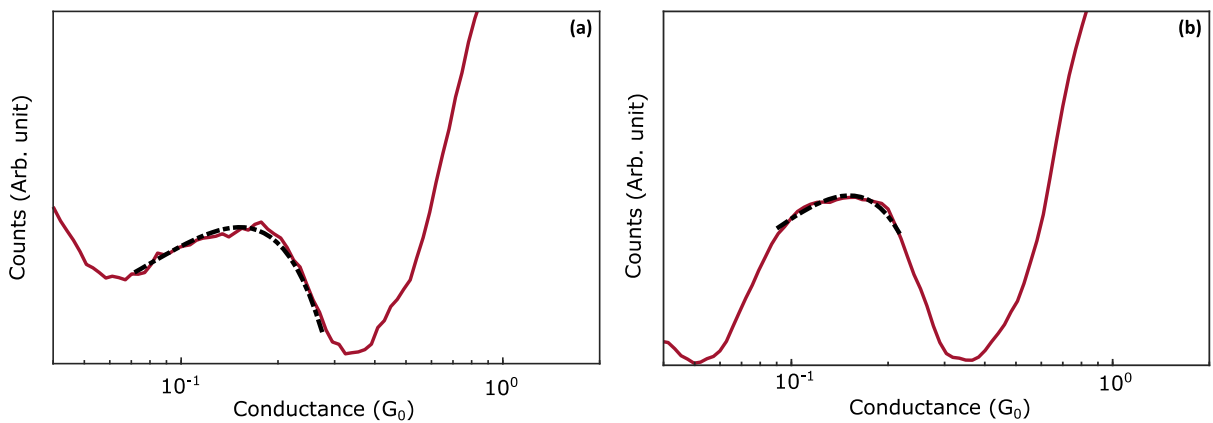

*Figure S6:* Conductance histogram of Au/ferrocene/Au junction in logarithmic scale for two different bias voltage 20mV (a) and 50mV (b) where black dash dot line represents the Gaussian fitting of the corresponding molecular peak with peak value $(1.40 \pm 0.02) \times 10^{-1}$ $G_0$ and $(1.60 \pm 0.01) \times 10^{-1}$ $G_0$.

**(c) Precursor configuration**

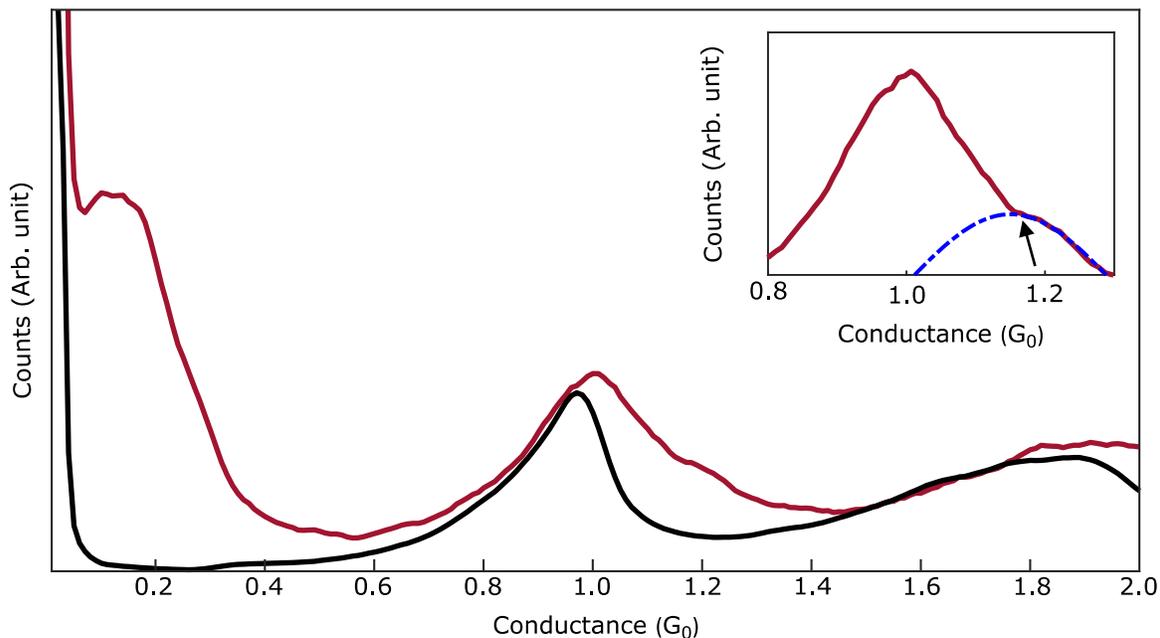

*Figure S7:* Conductance histogram constructed from the conductance traces with (red) and without (black) molecular features. Inset: Zoomed view of the same histogram where blue dash-dot line presents the Gaussian fitting of the corresponding precursor peak which yields a maxima at $1.18 \pm 0.02\ G_0$.

**(d) Conditional analysis**

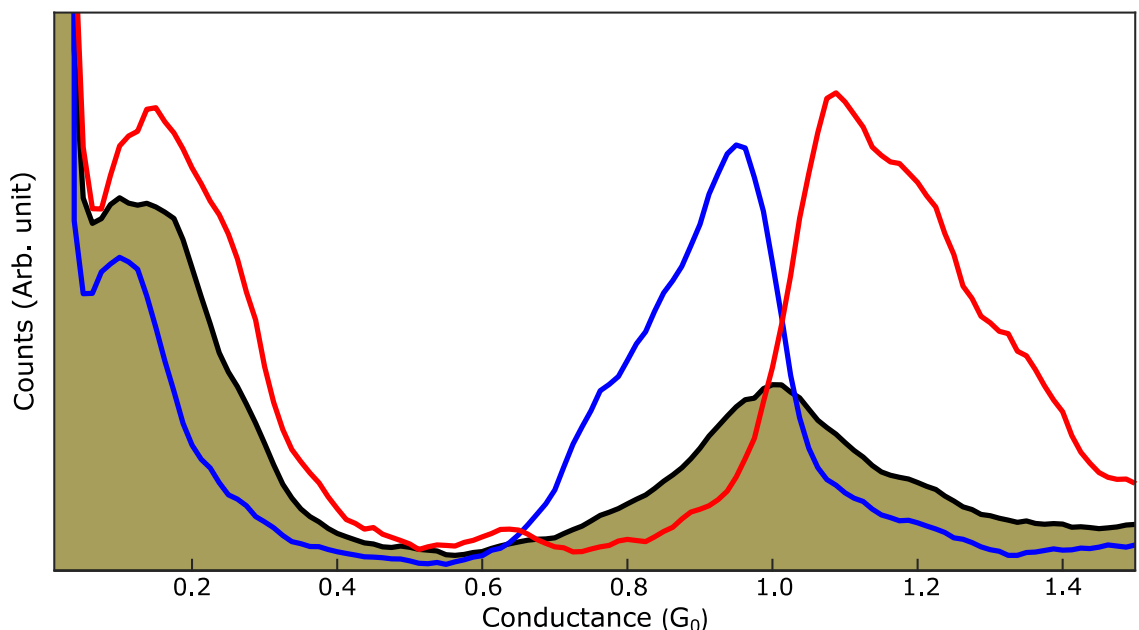

**Figure S8:** Conditional histograms for selected traces with larger than average length in different regions: R1- molecular conductance region (Red) and R2 - single atomic contacts region (Blue). As a reference the yellowish area graph shows the histogram for all traces.

## 4. Additional theoretical results
### (a) Binding energy curve

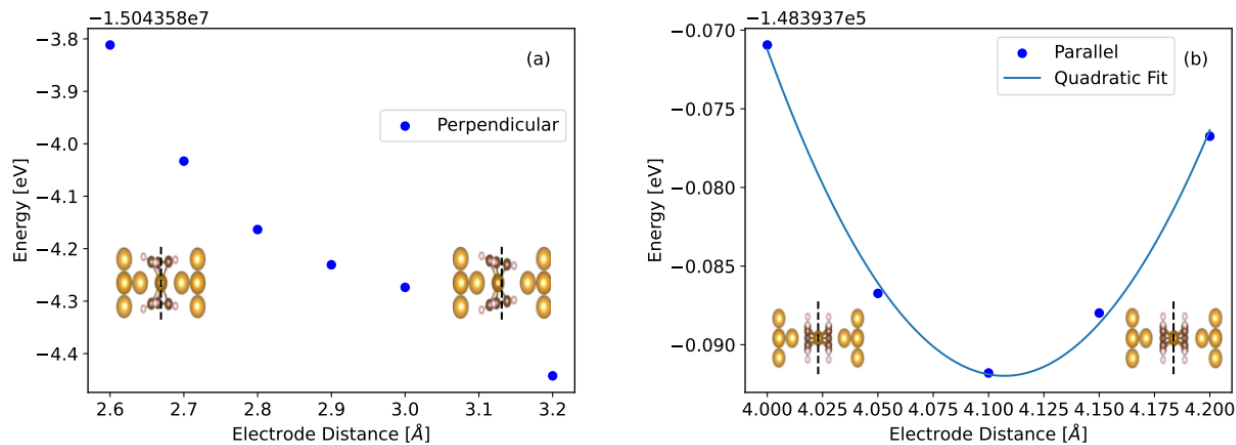

*Figure S9:* *(a-b) Energy of the extended molecule for different distances of the electrodes from the molecule. For perpendicular geometry (a), no clear minimum is observed for distances around 2.8 Å. However, the molecule occupies a symmetric junction position only for electrodes at this distance or closer. Asymmetric tilting of molecule towards one electrode is illustrated in detail of the junction geometry. For parallel geometry (b), the minimum is observed and the energy dependence is relatively weak. For these positions of electrodes, molecule always forms a symmetric junction.*

### (b) Molecular orbitals of free ferrocene

Ferrocene without presence of electrodes has HOMO and LUMO, with HOMO being the dominant transport state.

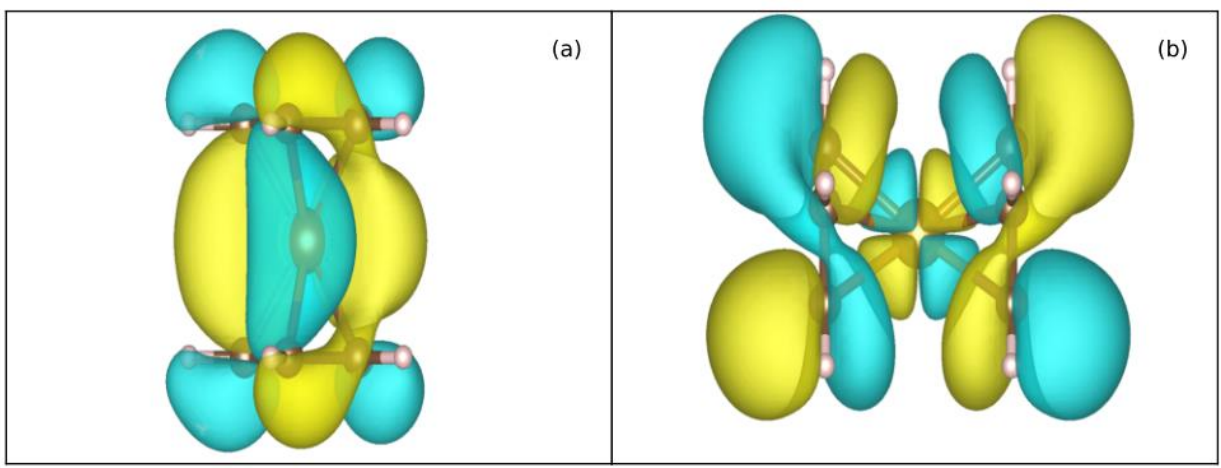

*Figure S10: (a-b) The highest occupied molecular orbital (HOMO) for free ferrocene (a) and lowest unoccupied molecular orbital (LUMO) for the same molecule (b). The LUMO state is present on the ferrocene extended molecule, as shown in figure 5a-b.*

### (c) Conductance of perpendicular and parallel geometry with applied voltage

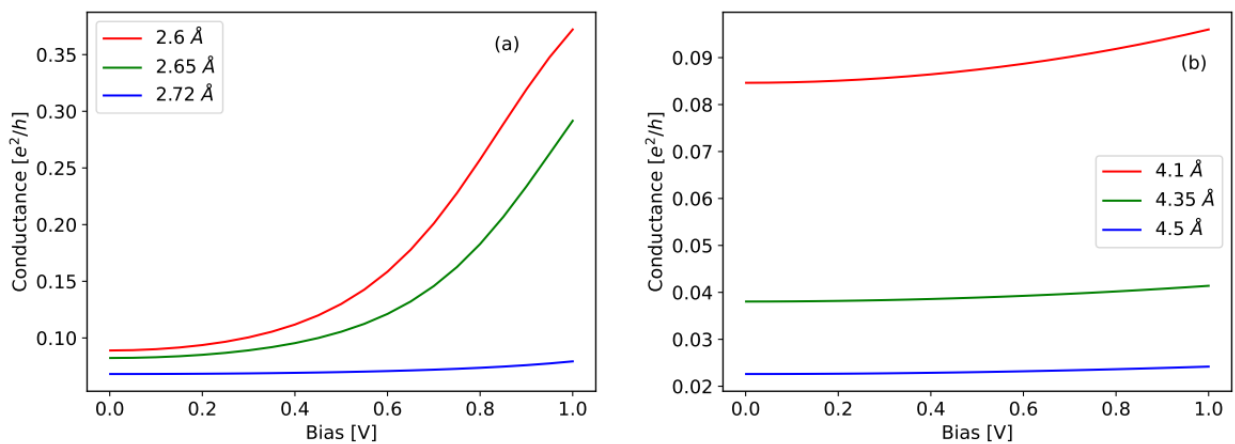

*Figure S11: (a-b) Conductance (nominal, i.e., current divided by voltage) calculated as integral of the transmission function with transport window kernel. For perpendicular geometry (a), the broadening of the transport window by increased bias voltage leads to significant conductance – the window now encompasses a transmission function peak. For parallel geometry (b), voltage applied would need to be larger to achieve the same conductance as in perpendicular geometry.*

**(d) Effect of ring rotation of Cp rings on transmission**

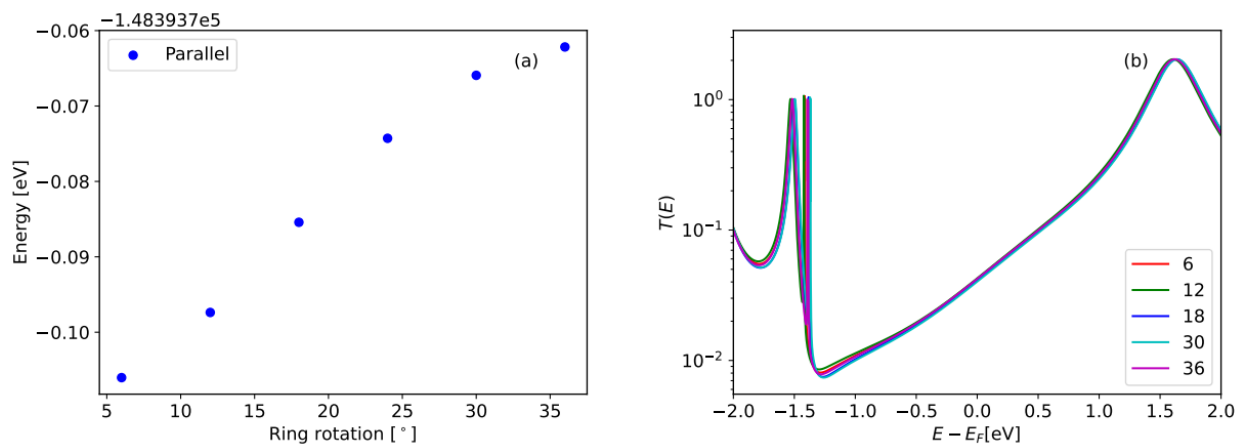

*Figure S12: (a) Energy of the extended molecule in parallel geometry for different relative rotations of the cyclopentacene rings (0° would represent eclipsed rings). Energy scale is about 40 meV, which is accessible at room temperature. However, the ring rotations have no discernible influence on the shape of transmission function. Hence, only a single relative position of the rings is considered in rest of the text.*

**(e) Transmission function of perpendicular and parallel geometry**

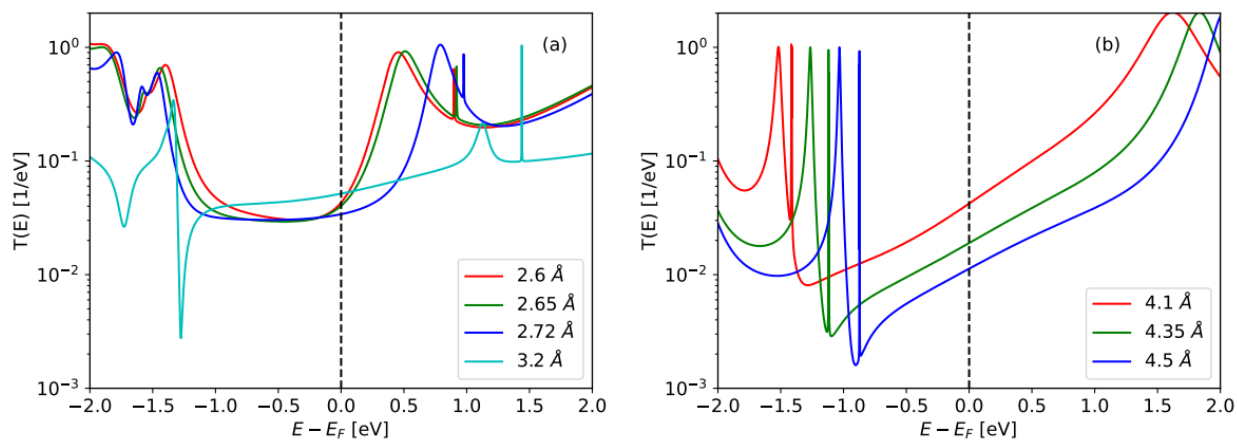

*Figure S13*: *(a-b) Transmission functions with varying distance of electrodes from the junction centre in perpendicular (a) and parallel (b) geometries of ferrocene molecular junction. For smaller distance, the transmission increases overall, as the overlap of the orbitals increases. Furthermore, peaks tend to shift towards lower energy as the distance decreases. Calculations for parallel geometry are done in TURBOMOLE, for perpendicular geometry, the calculations are done in FHI-AIMS with exception of calculation at 3.2 Å, which is also done in TURBOMOLE – hence wee see that main features of the transmission function are same in TURBOMOLE and in FHI-AIMS.*

**(f) Influence of Van der Waals correction and nature of adsorption**

We investigate the bond energy in the following way – we calculate the DFT total energy of molecule + electrodes system (extended molecule), $E_{//}$ for parallel geometry (at 4.1 Å) and $E_\perp$ for perpendicular geometry (at 2.8 Å) with Van der Waals/dispersion corrections. Then the energy of the same system without the central molecule (electrodes only) $E_{//\perp,E}$ and the energy of the ferrocene molecule alone (in gas phase), $E_g$ are subtracted, resulting in the estimate of bond energy. Same analysis is carried out for the system without Van der Waals/dispersion corrections, using energy of the extended molecule without dispersion $E_{//\perp,N}$, energy of electrodes without dispersion $E_{//\perp,E,N}$ and energy of gas phase ferrocene without dispersion $E_{g,N}$. Results are summarized in Table S1.

We can see that at the given distance, the perpendicular geometry is stabilised by Van der Waals interaction – without it, the bond is not energetically favourable to dissociated state.

| Name of the calculation | Energy, perpendicular ($x = \perp$) [eV] | Energy, parallel ($x=//$) [eV] |
|---|---|---|
| Bond energy with VdW ($E_x - E_{x,E} - E_g$) | -0.66 | -1.61 |
| Bond energy without VdW ($E_{x,N} - E_{x,E,N} - E_{g,N}$) | 0.11 | -0.52 |

*Table S1:* *Bond energy for the perpendicular and parallel geometries. The used electrode distance is 2.8 Å for perpendicular and 4.1 Å for parallel calculations. For perpendicular geometry, FHI-AIMS program was used, for parallel geometry, TURBOMOLE program was used.*

### (g) Local Density of States on the Iron Ion

The local density of states (LDOS) is calculated as a projection of spectral density operator onto the basis of atomic orbitals of a given atom/ion[9]. For a Green's function $G(E)$ in the basis of atomic orbitals, local density of states $\rho(E)$ for set of chosen orbitals $I$ is given as

$$\rho(E) = \frac{-1}{\pi} \sum_{i \in I} \Im\, G_{ii}(E)$$

In our case, since we claim that the transport occurs through the metal-metal bond in the perpendicular geometry, it is reasonable to investigate the LDOS on the iron ion at the centre of the ferrocene molecule. The calculation is carried out in the AITRANSS program. The resultant LDOS in both parallel and perpendicular geometry is shown in Fig. S14. Compare these results with the transmission curves presented in Fig. 5 in the main text – the features present in the transmission function are also present in the LDOS.

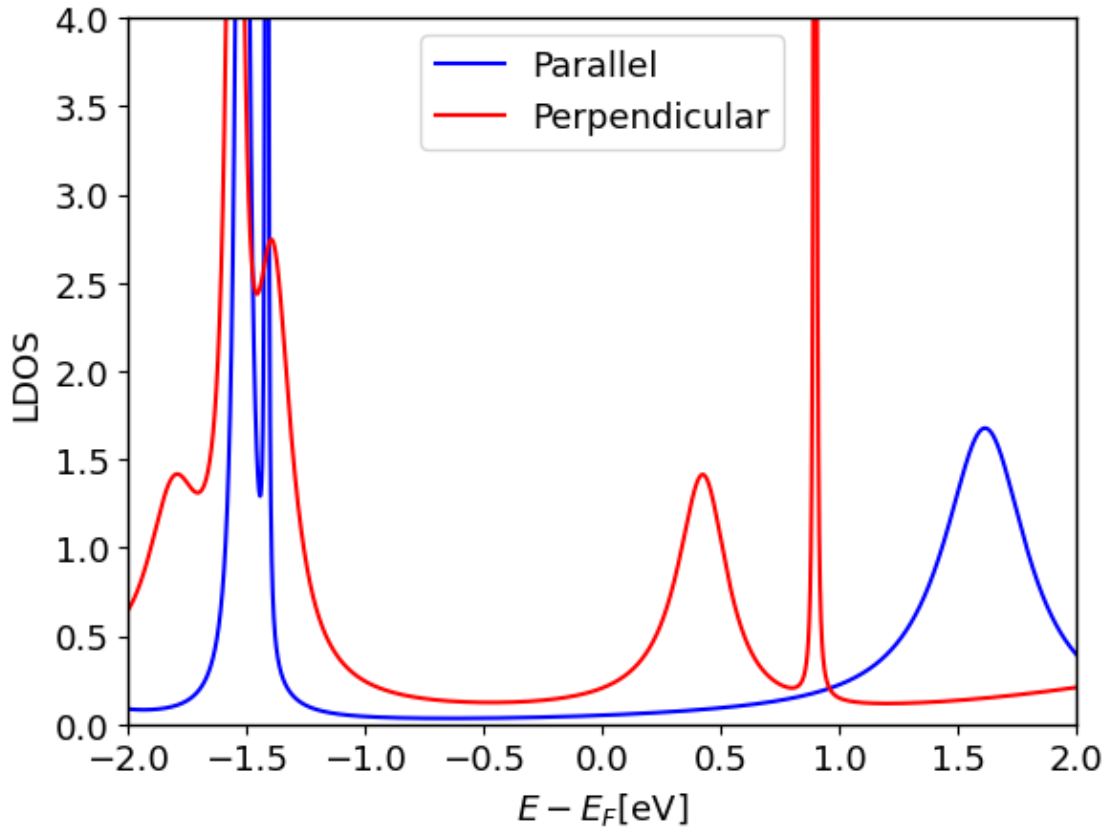

*Figure S14: The local density of states (LDOS) on the iron ion at the centre of the ferrocene molecule. In the parallel geometry, the LUMO transport state is further away from the Fermi energy than in the perpendicular geometry. The apex-to-iron distance were 2.6 Å in perpendicular geometry and 4.1 Å in parallel geometry. The contributions to transport are different from different states, but their location is similar to location of resonances in the transmission curve.*

## (h) Stability of Observables – Basis Set Size and Ad-Atoms

Two considerations should be taken into account when determining the convergence behaviour of the observables derived from the ab-initio calculations. Firstly, we should be certain that the basis sets used to represent the molecular orbitals are sufficiently large so that the states (especially the excited states) can be accurately modelled. Secondly, at room temperatures, the surface of the gold electrodes is not perfectly crystalline. We model this by repeating the calculations with randomly placed ad-atoms (see Fig. S16) that simulate deviations from the crystalline order. In Fig. S15, we can compare the results obtained for smaller basis set size and for the same basis set size but including add-atoms with the results obtained in the rest of the text. We can see that in both cases, the changes to both the transmission function and the local density of states are small, on the order of few percent of the magnitude of the respective observables.

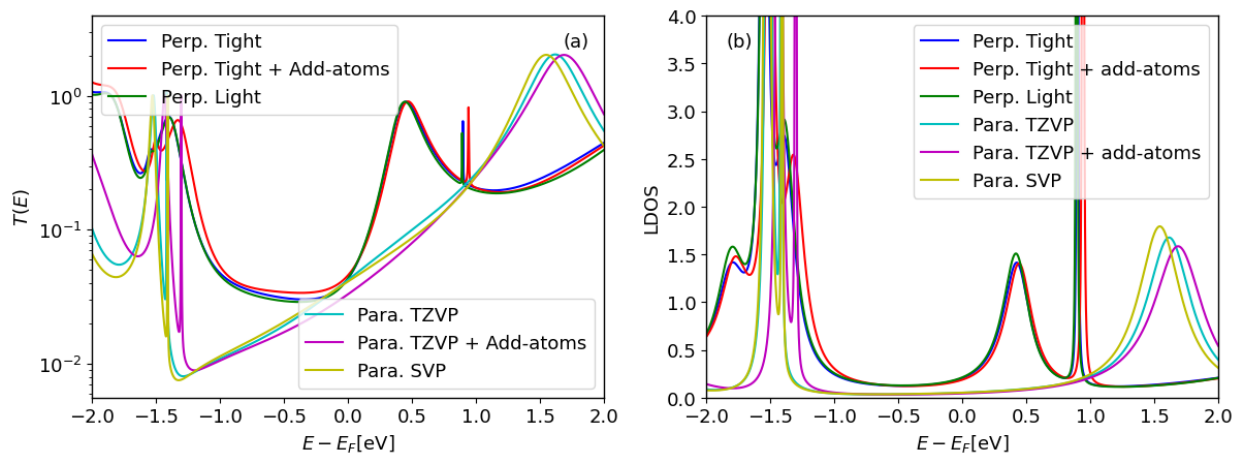

*Figure S15:* *The transmission function (a) and local density of states (b) on the iron ion at the centre of the ferrocene molecule for different basis sets and with the inclusion of add-atoms. Changes to both observables are small – on the order of few percent. The qualitative results of the ab-initio calculations are therefore numerically (basis) and configurationally (add-atoms) stable.*

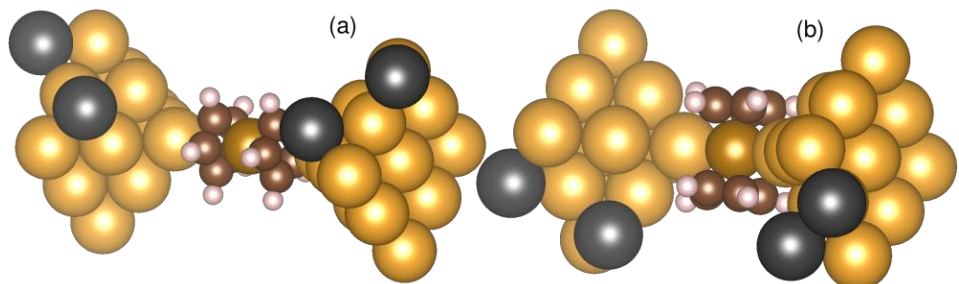

*Figure S16:* *The geometries of the parallel (a) and perpendicular (b) junctions with ad-atoms placed on the electrodes. Ad-atoms are gold atoms, here their color is changed to gray to explicitly differentiate them from the regular (crystalline) gold atoms.*

## (i) Calculation of transmission function constrained to a single scattering state

The transmission function is calculated in the Landauer-Buttiker approach in the AITRANSS package, as described in the main text. This involves derivation of the transmission function as trace of the matrix product of matrices based on NEGFs of the extended molecule. Dropping the (orbital) indices for certain orbitals allows us to see the transmission function without the influence of the orbitals with given indices. Alternatively, all indices except one can be set to zero, leaving us with a single peak, corresponding to non-interfering transport through single orbital. Numerically, this is done by including a masking matrix $M_{ij}$, which has entries either 1 or 0, with 1 only at indices with orbitals to be considered for transmission, i.e.

$$T = \sum_{i,j} M_{ij}\, \Gamma_{L,ij} G_j \Gamma_{R,ji} G_i^c$$

where the symbols used have the same meaning as given in the AITRANSS package article[8] and the summation runs over indices of the non-equilibrium eigen states.

### (j) Coordinates of Converged Junctions

We are including the xyz coordinate files of the junctions used for conductance and LDOS calculations in main text. For the perpendicular geometry of Ferrocene (FC) at distance 2.6 Å

```
AU 2.6 0.0 0.0
AU 4.635 2.035 0.0
AU 4.635 0.0 -2.035
AU 4.635 -2.035 0.0
AU 4.635 0.0 2.035
AU 6.67 4.07 0.0
AU 6.67 2.035 -2.035
AU 6.67 0.0 0.0
AU 6.67 2.035 2.035
AU 6.67 0.0 -4.07
AU 6.67 -2.035 -2.035
AU 6.67 -4.07 0.0
AU 6.67 -2.035 2.035
AU 6.67 0.0 4.07
C 0.97965951 0.72286369 -1.83333102
C 0.97965276 -0.72284304 -1.8333394
C -0.36393193 -1.16133116 -1.73942802
C -1.21474473 2.06e-05 -1.81068501
C -0.36392134 1.16136335 -1.73941294
H 1.83741061 1.36962039 -2.01043626
H 1.83739862 -1.36960559 -2.01044833
H -0.70582592 -2.19099436 -1.75493551
H -2.21956146 2.878e-05 -2.2709147
H -0.70580557 2.19102994 -1.75490815
FE 0.01542159 4.98e-06 3e-08
C -0.36392358 -1.1613507 1.73941577
C -1.21474471 -6.06e-06 1.81068505
C -0.36392968 1.16134382 1.73942527
C 0.97965417 0.72285316 1.83333784
C 0.97965812 -0.72285356 1.83333265
H -0.70580987 -2.19101658 1.74491323
H -2.21956146 -1.088e-05 2.2709147
H -0.7058216 2.19100772 1.75493052
H 1.83740143 1.36961424 2.01044514
H 1.83740782 -1.36961173 2.01043951
AU -2.6 0.0 0.0
AU -4.635 2.035 0.0
AU -4.635 0.0 2.035
AU -4.635 -2.035 0.0
AU -4.635 0.0 -2.035
AU -6.67 4.07 0.0
AU -6.67 2.035 2.035
AU -6.67 0.0 0.0
AU -6.67 2.035 -2.035
AU -6.67 0.0 4.07
AU -6.67 -2.035 2.035
AU -6.67 -4.07 0.0
AU -6.67 -2.035 -2.035
AU -6.67 0.0 -4.07
```

For the parallel geometry of ferrocene (FC) at distance 4.1 Å

```
49

AU 4.1000000000000005 0.0 0.0
AU 6.1349999999999945 2.034999999999999 0.0
AU 6.1349999999999945 0.0 -2.034999999999999
AU 6.1349999999999945 -2.034999999999999 0.0
AU 6.1349999999999945 0.0 2.034999999999999
AU 8.170000000000003 4.070000000000003 0.0
AU 8.170000000000003 2.034999999999999 -2.034999999999999
AU 8.170000000000003 0.0 0.0
AU 8.170000000000003 2.034999999999999 2.034999999999999
AU 8.170000000000003 0.0 -4.070000000000003
AU 8.170000000000003 -2.034999999999999 -2.034999999999999
AU 8.170000000000003 -4.070000000000003 0.0
AU 8.170000000000003 -2.034999999999999 2.034999999999999
AU 8.170000000000003 0.0 4.070000000000003
C -1.6275884823423405 1.1672011660639403 -0.3794940697340722
C -1.6280444640169234 1.9040681679270106e-06 -1.2275172039221152
C -1.6275938041695084 -1.1672002906265961 -0.37949638933260704
C -1.6272869449855327 -0.7214323180761611 0.992844894170462
C -1.6272841432496512 0.721430617767112 0.9928453696556596
H -1.6230710629197664 2.1995178676132316 -0.7146906088067383
H -1.6232334781413114 -9.218863237893784e-07 -2.312954673714635
H -1.6230798843302994 -2.1995158602355698 -0.7146947432813314
H -1.6234780155466357 -1.3589602783781292 1.8714272764475401
H -1.6234720284592095 1.358957808092817 1.8714292702044983
FE 6.0090768477092287e-05 -4.036063279545556e-06 -0.00023053059714474308
C 1.627267917833809 -0.7214326978809809 0.9928470583912591
C 1.6272658448969644 0.7214300848318275 0.9928473930260995
C 1.6275718834970363 1.1672065716085678 -0.37949628146332515
C 1.6280369178532152 1.3411948043539458e-06 -1.2275256323366734
C 1.6275771453266477 -1.1672061010152217 -0.3794988757873471
H 1.6234914534695386 -1.3590035157746307 1.8714880936174583
H 1.6234862036312574 1.3590002848383538 1.8714900456060093
H 1.6230865555481842 2.1995915927532743 -0.7147139486369102
H 1.6232514060329768 -1.3748423732642777e-06 -2.3130380071454466
H 1.6230957422834067 -2.1995902276311665 -0.7147183111014612
AU -4.1000000000000005 0.0 0.0
AU -6.1349999999999945 2.034999999999999 0.0
AU -6.1349999999999945 0.0 2.034999999999999
AU -6.1349999999999945 -2.034999999999999 0.0
AU -6.1349999999999945 0.0 -2.034999999999999
AU -8.170000000000003 4.070000000000003 0.0
AU -8.170000000000003 2.034999999999999 2.034999999999999
AU -8.170000000000003 0.0 0.0
AU -8.170000000000003 2.034999999999999 -2.034999999999999
AU -8.170000000000003 0.0 4.070000000000003
AU -8.170000000000003 -2.034999999999999 2.034999999999999
AU -8.170000000000003 -4.070000000000003 0.0
AU -8.170000000000003 -2.034999999999999 -2.034999999999999
AU -8.170000000000003 0.0 -4.070000000000003
```

For anchoring group coupled 1,1′-bis(aminomethyl)ferrocene (FC-NH$_2$)

```
59

H    6.925502e+00  4.942298e+00  6.607842e+00
H    6.980205e+00  2.650830e+00  7.374509e+00
N    6.507916e+00  4.675867e+00  7.512306e+00
C    7.191464e+00  3.499583e+00  8.058416e+00
H    6.612877e+00  5.490762e+00  8.134631e+00
H    9.100405e+00  5.816601e+00  8.231419e+00
C    8.692103e+00  3.593483e+00  8.241950e+00
C    9.484256e+00  4.794415e+00  8.257599e+00
C    1.087128e+01  4.426156e+00  8.311816e+00
C    9.604039e+00  2.483432e+00  8.314512e+00
H    9.327490e+00  1.426561e+00  8.329731e+00
H    1.172283e+01  5.109442e+00  8.345576e+00
C    1.094514e+01  2.994194e+00  8.348136e+00
H    1.185701e+01  2.397103e+00  8.405771e+00
H    6.701868e+00  3.231766e+00  9.013912e+00
Fe   9.889768e+00  3.693187e+00  9.940453e+00
H    6.213784e+00  3.897110e+00  1.085850e+01
H    7.340496e+00  1.749985e+00  1.116580e+01
H    1.109847e+01  5.741479e+00  1.145433e+01
H    1.220093e+01  3.259619e+00  1.148205e+01
C    1.055004e+01  4.799452e+00  1.152599e+01
C    1.113528e+01  3.489383e+00  1.154051e+01
H    8.389237e+00  5.453417e+00  1.157986e+01
C    9.122281e+00  4.643879e+00  1.159113e+01
C    1.006820e+01  2.531723e+00  1.160680e+01
H    1.018658e+01  1.446202e+00  1.162325e+01
C    8.816670e+00  3.240499e+00  1.165267e+01
N    6.279121e+00  3.447924e+00  1.178355e+01
C    7.467251e+00  2.597673e+00  1.186954e+01
H    6.324841e+00  4.204410e+00  1.248176e+01
H    7.466900e+00  2.137877e+00  1.288214e+01
Au   6.377095e+00  3.697642e+00  5.526543e+00
Au   7.816057e+00  2.258680e+00  3.491543e+00
Au   7.816057e+00  5.136604e+00  3.491543e+00
Au   4.938133e+00  5.136604e+00  3.491543e+00
Au   4.938133e+00  2.258680e+00  3.491543e+00
Au   9.255020e+00  8.197173e-01  1.456543e+00
Au   9.255020e+00  3.697642e+00  1.456543e+00
Au   6.377095e+00  3.697642e+00  1.456543e+00
Au   6.377095e+00  8.197173e-01  1.456543e+00
Au   9.255020e+00  6.575566e+00  1.456543e+00
Au   6.377095e+00  6.575566e+00  1.456543e+00
Au   3.499171e+00  6.575566e+00  1.456543e+00
Au   3.499171e+00  3.697642e+00  1.456543e+00
Au   3.499171e+00  8.197173e-01  1.456543e+00
Au   6.377095e+00  3.697642e+00  1.376904e+01
Au   7.816057e+00  2.258680e+00  1.580404e+01
Au   7.816057e+00  5.136604e+00  1.580404e+01
Au   4.938133e+00  5.136604e+00  1.580404e+01
Au   4.938133e+00  2.258680e+00  1.580404e+01
Au   9.255020e+00  8.197173e-01  1.783904e+01
Au   9.255020e+00  3.697642e+00  1.783904e+01
Au   6.377095e+00  3.697642e+00  1.783904e+01
Au   6.377095e+00  8.197173e-01  1.783904e+01
Au   9.255020e+00  6.575566e+00  1.783904e+01
Au   6.377095e+00  6.575566e+00  1.783904e+01
Au   3.499171e+00  6.575566e+00  1.783904e+01
Au   3.499171e+00  3.697642e+00  1.783904e+01
Au   3.499171e+00  8.197173e-01  1.783904e+01
```

For anchoring group coupled 1,1′-dicyanoferrocene (FC-CN)

51

```
N    5.404041e+00  2.996933e+00  1.142563e+01
C    6.558563e+00  3.204079e+00  1.148250e+01
C    7.953150e+00  3.472143e+00  1.151889e+01
H    8.002465e+00  5.730986e+00  1.152239e+01
C    8.550918e+00  4.788277e+00  1.153238e+01
C    9.970659e+00  4.614990e+00  1.153808e+01
C    9.017297e+00  2.492573e+00  1.154076e+01
H    8.883532e+00  1.409709e+00  1.154121e+01
C    1.025784e+01  3.208797e+00  1.154291e+01
H    1.070626e+01  5.419941e+00  1.156112e+01
H    1.125240e+01  2.761924e+00  1.157012e+01
Fe   9.119944e+00  3.715408e+00  1.317074e+01
H    7.944199e+00  5.714581e+00  1.479115e+01
C    8.507333e+00  4.781622e+00  1.479786e+01
C    9.000109e+00  2.493293e+00  1.480037e+01
H    8.866322e+00  1.410988e+00  1.480060e+01
C    7.925030e+00  3.458802e+00  1.480322e+01
H    1.065847e+01  5.437564e+00  1.480905e+01
H    1.123077e+01  2.778390e+00  1.481242e+01
C    9.931441e+00  4.624671e+00  1.481414e+01
C    1.023332e+01  3.220722e+00  1.481572e+01
C    6.537727e+00  3.151517e+00  1.482012e+01
N    5.394238e+00  2.887504e+00  1.486477e+01
Au   5.083522e+00  3.184248e+00  9.412057e+00
Au   3.644559e+00  1.745285e+00  7.377057e+00
Au   6.522484e+00  1.745285e+00  7.377057e+00
Au   6.522484e+00  4.623210e+00  7.377057e+00
Au   3.644559e+00  4.623210e+00  7.377057e+00
Au   2.205597e+00  3.063232e-01  5.342057e+00
Au   5.083522e+00  3.063232e-01  5.342057e+00
Au   5.083522e+00  3.184248e+00  5.342057e+00
Au   2.205597e+00  3.184248e+00  5.342057e+00
Au   7.961446e+00  3.063232e-01  5.342057e+00
Au   7.961446e+00  3.184248e+00  5.342057e+00
Au   7.961446e+00  6.062172e+00  5.342057e+00
Au   5.083522e+00  6.062172e+00  5.342057e+00
Au   2.205597e+00  6.062172e+00  5.342057e+00
Au   5.083522e+00  3.184248e+00  1.687838e+01
Au   3.644559e+00  1.745285e+00  1.891338e+01
Au   6.522484e+00  1.745285e+00  1.891338e+01
Au   6.522484e+00  4.623210e+00  1.891338e+01
Au   3.644559e+00  4.623210e+00  1.891338e+01
Au   2.205597e+00  3.063232e-01  2.094838e+01
Au   5.083522e+00  3.063232e-01  2.094838e+01
Au   5.083522e+00  3.184248e+00  2.094838e+01
Au   2.205597e+00  3.184248e+00  2.094838e+01
Au   7.961446e+00  3.063232e-01  2.094838e+01
Au   7.961446e+00  3.184248e+00  2.094838e+01
Au   7.961446e+00  6.062172e+00  2.094838e+01
Au   5.083522e+00  6.062172e+00  2.094838e+01
Au   2.205597e+00  6.062172e+00  2.094838e+01
```

**(k) Binding energy of anchoring group coupled FC-NH$_2$ and FC-CN**

*Table S2: Binding energy of FC-NH$_2$ and FC-CN*

| Molecular system | Binding energy of Anchoring geometry |
|---|---|
| 1,1′-bis(aminomethyl)ferrocene (FC-NH$_2$) | -1.20 eV |
| 1,1′-dicyanoferrocene (FC-CN) | -1.13 eV |

**References:**

(1) Teresa González, M.; Wu, S.; Huber, R.; Van Der Molen, S. J.; Schönenberger, C.; Calame, M. Electrical Conductance of Molecular Junctions by a Robust Statistical Analysis. *Nano Lett* **2006**, *6* (10), 2238–2242. https://doi.org/10.1021/nl061581e.

(2) Martin, C. A.; Ding, D.; Sørensen, J. K.; Bjørnholm, T.; Van Ruitenbeek, J. M.; Van Der Zant, H. S. J. Fullerene-Based Anchoring Groups for Molecular Electronics. *J Am Chem Soc* **2008**, *130* (40), 13198–13199. https://doi.org/10.1021/ja804699a.

(3) Quek, S. Y.; Kamenetska, M.; Steigerwald, M. L.; Choi, H. J.; Louie, S. G.; Hybertsen, M. S.; Neaton, J. B.; Venkataraman, L. Mechanically Controlled Binary Conductance Switching of a Single-Molecule Junction. *Nat Nanotechnol* **2009**, *4* (4), 230–234. https://doi.org/10.1038/nnano.2009.10.

(4) Chen, F.; Li, X.; Hihath, J.; Huang, Z.; Tao, N. Effect of Anchoring Groups on Single-Molecule Conductance: Comparative Study of Thiol-, Amine-, and Carboxylic-Acid-Terminated Molecules. *J Am Chem Soc* **2006**, *128* (49), 15874–15881. https://doi.org/10.1021/ja065864k.

(5) Hong, W.; Manrique, D. Z.; Moreno-García, P.; Gulcur, M.; Mishchenko, A.; Lambert, C. J.; Bryce, M. R.; Wandlowski, T. Single Molecular Conductance of Tolanes: Experimental and Theoretical Study on the Junction Evolution Dependent on the Anchoring Group. *J Am Chem Soc* **2012**, *134* (4), 2292–2304. https://doi.org/10.1021/ja209844r.

(6) Makk, P.; Tomaszewski, D.; Martinek, J.; Balogh, Z.; Csonka, S.; Wawrzyniak, M.; Frei, M.; Venkataraman, L.; Halbritter, A. Correlation Analysis of Atomic and Single-Molecule Junction Conductance. *ACS Nano* **2012**, *6* (4), 3411–3423. https://doi.org/10.1021/nn300440f.

(7) Perrin, M. L.; Martin, C. A.; Prins, F.; Shaikh, A. J.; Eelkema, R.; van Esch, J. H.; van Ruitenbeek, J. M.; van der Zant, H. S. J.; Dulić, D. Charge Transport in a Zinc-Porphyrin Single-Molecule Junction. *Beilstein J Nanotechnol* **2011**, *2* (1), 714–719. https://doi.org/10.3762/bjnano.2.77.

(8) Arnold, A.; Weigend, F.; Evers, F. Quantum Chemistry Calculations for Molecules Coupled to Reservoirs: Formalism, Implementation, and Application to Benzenedithiol. *J Chem Phys* **2007**, *126* (17). https://doi.org/10.1063/1.2716664.

(9) Bagrets, A. Spin-polarized electron transport across metal–organic molecules: a density functional theory approach. *J. Chem. Theory Comput.* **2013**, *9*, 2801–2815.